\documentclass[twocolumn,showpacs,amsmath,amssymb,prd,showkeys,10pt,nofootinbib,superscriptaddress]{revtex4}

\usepackage{epsfig}
\usepackage{latexsym}
\usepackage{graphicx}
\usepackage{amsfonts}
\usepackage{amsmath}
\usepackage{natbib}
\usepackage{xcolor}
\usepackage{wasysym}
\usepackage{bm}
\usepackage{array}

\newcommand{\etal}{{et~al.}}
\newcommand{\sa}{\mathrm{S}}
\newcommand{\te}{\mathrm{T}}
\newcommand{\pk}{\mathcal{P}}

\def\apj{Astrophys. J.}%
\def\aap{Astron. \& Astrophys.}%
\def\prd{Phys.~Rev.~D}%
\def\prl{Phys.~Rev.~Lett.}%
\def\ds{\displaystyle}

\def\simlt{\lower.5ex\hbox{$\; \buildrel < \over \sim \;$}}
\def\simgt{\lower.5ex\hbox{$\; \buildrel > \over \sim \;$}}

\begin{document}
\title{Detecting chiral gravity with the pure pseudospectrum reconstruction of the cosmic microwave background polarized anisotropies}

\author{ A. Fert\'e}
\email{agnes.ferte@ias.u-psud.fr}
\affiliation{%
 Universit\'e Paris-Sud 11, Institut d'Astrophysique Spatiale, UMR8617, Orsay, France, F-91405}
\affiliation{%
CNRS, Orsay, France, F-91405}
\email{agnes.ferte@ias.u-psud.fr}

\author{J. Grain}
 \email{julien.grain@ias.u-psud.fr}
\affiliation{%
CNRS, Orsay, France, F-91405}
 \affiliation{%
 Universit\'e Paris-Sud 11, Institut d'Astrophysique Spatiale, UMR8617, Orsay, France, F-91405}
\email{julien.grain@ias.u-psud.fr}

\begin{abstract}
We consider the possible detection of parity violation at the linear level in gravity using polarized anisotropies of the cosmic microwave background. Since such a parity violation would lead to non-zero $TB$ and $EB$ correlations, this makes those {\it odd-parity} angular power spectra a potential probe of parity violation in the gravitational sector. These spectra are modeled incorporating the impact of lensing and we explore their possible detection in the context of small-scale (balloon-borne or ground-based) experiments and a future satellite mission dedicated to $B$-mode detection. We assess the statistical uncertainties on their reconstruction using mode-counting and a (more realistic) pure pseudospectrum estimator approach. Those uncertainties are then translated into constraints on the level of parity asymmetry. We found that detecting chiral gravity is impossible for ongoing small-scale experiments. However, for a satellite-like mission, a parity asymmetry of 50\% could be detected at 68\% of confidence level (at least, depending on the value of the tensor-to-scalar ratio), and a parity asymmetry of 100\% is measurable with {\it at least} a confidence level of 95\%. We also assess the impact of a possible miscalibration of the orientation of the polarized detectors, leading to spurious $TB$ and $EB$ cross-correlations. We show that in the context of pseudospectrum estimation of the angular power spectra, self-calibration of this angle could significantly reduce the statistical significance of the measured level of parity asymmetry (by {\it e.g.} a factor $\sim2.4$ for a miscalibration angle of 1 degree). For chiral gravity and assuming a satellite mission dedicated to primordial $B$-mode, a non detection of the $TB$ and $EB$ correlation would translate into an upper bound on parity violation of 39\% at 95\% confidence level for a tensor-to-scalar ratio of 0.2, excluding values of the (imaginary) Barbero-Immirzi parameter comprised between 0.2 and 4.9 at 95\% CL.
\end{abstract}


\maketitle

\section{Introduction}
\label{sec:intro}

The anisotropies of the cosmic microwave background (CMB) are currently the most powerful probe of the physics underlying the primordial universe. Those anisotropies arise into three flavors: total intensity and two degrees of freedom describing its linear polarization. Though CMB polarized anisotropies are measured using two Stokes parameters, $Q$ and $U$, they are most conveniently described using a gradient-like, $E$-mode, and a curl-like, $B$-mode \cite{zaldarriaga_seljak_1997,kamionkowski_etal_1997}. 

Such a decomposition of the linearly polarized anisotropies is meaningful at a physical level as it is directly linked to the primordial cosmological perturbations sourcing CMB anisotropies. For instance, on the linear level the $B$-modes can be sourced by the primordial gravitational waves \cite{seljak_zaldarriaga_1997,spergel_zaldarriaga_1997} and not by the scalar fluctuations, thought to be largely responsible for the observed total intensity and $E$-mode anisotropies. Consequently, a detection of the $B$-mode anisotropy at large angular scales ($\ell \apprle 100$) in excess of what is expected from the gravitational lensing signal could be seen as a direct validation of inflationary theories, as the latter are considered to be the most likely source of the gravity waves, and could allow for discrimination between different inflationary models. It could also set useful constraints on the reionization period \cite{zaldarriaga_1997}. At smaller angular scales, $B$-modes are expected to be mainly due to gravitational lensing of CMB photons which converts $E$-modes into $B$-modes \cite{zaldarriaga_seljak_1998} and therefore their detection -- a source of constraints on the matter perturbation evolution at redshift $z\sim1$ when light massive neutrinos and elusive dark energy both play potentially visible roles. Very recently, the {\it direct} detection of the lensing-induced $B$-mode and the primordial $B$-mode has been reported by the {\sc polarbear} experiment \cite{polarbearpaper} and the {\sc bicep2} experiment \cite{biceppaper}, respectively.

In the standard cosmological paradigm, the $TB$ end $EB$ cross-correlations are vanishing. However, they remain important quantities to be estimated from the data. This is because on the one hand, these odd-parity cross-spectra are comprehensive, end-to-end, null tests of the presence of instrumental and/or astrophysical systematic effects still present in the data (see {\it e.g.} \cite{hu_etal_2003,yadav_etal_2010}). On the other hand, as some non-standard cosmological mechanisms could produce nonvanishing odd-parity cross-spectra, their detection
could become a smoking gun of such effects with potentially far-reaching consequences for our understanding of the Universe. Examples of such mechanisms include a primordial stochastic magnetic field,
 which generates $TB$ and $EB$ correlations, if this magnetic field possesses a helical component \cite{pogosian_etal_2002,caprini_etal_2004,kahniashvili_etal_2005} or a pseudoscalar inflaton field which naturally couples to the electromagnetic field in a parity-dependant way \cite{sorbo_2011,anber_sorbo,cook_sorbo}. Similar effects can be obtained due to a rotation of the plane of linear polarization of the  CMB photons traveling from the last scattering surface to our detectors. This could result from either the Faraday rotation induced by interaction with background magnetic fields \cite{kosowsky_loeb_1996,kosowsky_etal_2005,campanelli_etal_2004,scoccola_etal_2004} or interactions with pseudoscalar fields on the trajectory of CMB photons \cite{carroll_1998}. 

In this paper, we consider the case of odd-parity angular power spectra as probes of parity violation in the primordial Universe as induced by gravity. The implication of chiral gravity on CMB anisotropies has been first explored in Ref. \cite{lue_etal_1999}  and then in Ref. \cite{contaldi_etal_2008} where it was shown that if parity is violated by gravitation at the linear level, CMB polarized anisotropies should exhibit non vanishing $EB$ and $TB$ cross-correlations.  This idea has been theoretically strengthened in Refs. \cite{magueijo_benincasa_2011,bethke_magueijo_2011a,bethke_magueijo_2011b}, and the idea that gravity could be parity dependent can be traced back to its formulation by {\it e.g.} Cartan and Kibble \cite{kibble_1961} or Ashtekar \cite{ashtekar_1986}. The possible detection of such parity asymmetry using CMB datas coming from a satellite-like mission has been discussed in Refs. \cite{xia_2012,saito_2007}, in Ref. \cite{wang_etal_2013} in the Horava-Lifshitz framework and in Ref. \cite{gluscevic_etal_2010} (including the case of a ballon-borne experiment in the latter).

We amend and elaborate on this proposal of Refs.\cite{lue_etal_1999,contaldi_etal_2008,xia_2012,saito_2007,gluscevic_etal_2010} in three directions. First, chiral gravity leads to {\it primary} $TB$ and $EB$ cross-correlations which are latter on, deformed by the weak gravitational lensing by large scale structure. As this could potentially lead {\it e.g.} $EE$ correlations to leak into $EB$ correlations (which would partially mask the primary $EB$), we therefore include in the predicted $C_\ell$'s the impact of lensing. Second, we make use of a Fisher matrix formalism to assess the potential detection of chiral gravity from the measurements of CMB polarized anisotropies in two typical experimental setups: small-scale experiments as motivated by operating (or forthcoming) balloon-borne or ground-based experiments such as \textsc{polarbear}, \textsc{sptpol}, {\sc qubic} or \textsc{actpol}, for ground-based experiments \cite{ground}, and, such as {\sc spider} or \textsc{ebex}, for balloon-borne experiments \cite{balloon}, and, satellite-like missions as motivated by {\it e.g.} {\sc l}ite\textsc{bird}, \textsc{prism} or \textsc{pix}i\textsc{e} proposals \cite{satellite}. Estimation of the uncertainties on the reconstructed $C^{TB(EB)}_\ell$ (subsequently used in the Fisher matrix) is based first on a na\"\i{ve} mode-counting (as a reference), and, second, on Monte-Carlo simulations coupled to a {\it realistic} statistical, pure pseudospectrum based estimators of angular power spectra. Thirdly, we assess the impact of a miscalibration of the orientation of the polarized detectors which creates spurious $TB$ and $EB$ correlations coming from $TE$ and $EE,~BB$ respectively.

The paper is organized as follows. The section \ref{sec:cell} is devoted to the theoretical prediction of the $TB$ and $EB$ angular power spectra including the impact of weak gravitational lensing by large scale structure. We present the statistical uncertainties on the reconstruction of $C^{TB(EB)}_\ell$'s using pure pseudospectrum estimators in Sec. \ref{sec:uncer}. The results of the application of such an approach to the two above-defined typical cases of CMB experiments dedicated to polarization, small-scale experiments and satellite-like missions, are presented in Secs. \ref{sec:satres} and \ref{sec:smares} respectively. We finally conclude and discuss the potential detection of chiral gravity within CMB anisotropies in the last section, Sec. \ref{sec:concl}, and discuss the relevance and extension of those results to other possible sources of parity violation in the primordial universe.

The technical details are provided in the appendices \ref{app:lensing} and \ref{appb}.

\section{Angular power spectra in chiral gravity}
\label{sec:cell}
\subsection{Primary anisotropies}
If parity invariance is broken by gravity, the amount of gravitational waves produced during inflation differs from one helicity state to another. As a consequence, the {\it primary} CMB polarized anisotropies gain non vanishing $TB$ and $EB$ cross-correlations. Using the line of sight solution of the Boltzmann equation \cite{zaldarriaga_harari_1995} and following Ref. \cite{contaldi_etal_2008}, the different angular power spectra are given by
\begin{eqnarray}
	C^{XZ}_\ell&=&\displaystyle\int dk \left\{\Delta^{X}_{\ell,\sa}(k,\eta_0)\Delta^{Z}_{\ell,\sa}(k,\eta_0)\pk_\sa(k)\right. \\
	&+&\left.\Delta^{X}_{\ell,\te}(k,\eta_0)\Delta^{Z}_{\ell,\te}(k,\eta_0)\left[\pk^R_\te(k)+\varepsilon\times\pk^L_\te(k)\right]\right\} \nonumber
\end{eqnarray}
In the above, $X,~Z=T,~E$ or $B$ and $\Delta^{X,\sa(\te)}_\ell$ is the transfer function for scalar(tensor) modes. The number $\varepsilon$ is equal to $(+1)$ for the $TT,~EE,~BB$ and $TE$ angular power spectra, and, equal to $(-1)$ for the $TB$ and $EB$ angular power spectra. Clearly, the $TB$ and $EB$ cross-correlations are equal to zero if $\pk^R_\te=\pk^L_\te$ at all $k$ values, as expected in a parity invariant primordial universe. However, if for any reason $\pk^R_\te(k)\neq \pk^L_\te(k)$, then the primary CMB anisotropies would exhibit non-vanishing $TB$ and $EB$ cross-correlations. In the following, the $TT,~EE,~BB$ and $TE$ correlations will be denoted {\it even} power spectra and the $TB$ and $EB$ correlations will be called {\it odd} power spectra.

The primary correlations of $BB$, $TB$ and $EB$ types are only sourced by the tensor mode and these angular power spectra are given by
\begin{eqnarray}
	C^{BB}_\ell&=&\displaystyle\int dk \left(\Delta^{B}_{\ell,,\te}(k,\eta_0)\right)^2\pk^{(+)}_\te(k), \\
	C^{TB}_\ell&=&\displaystyle\int dk \Delta^{T}_{\ell,\te}(k,\eta_0)\Delta^{B}_{\ell,\te}(k,\eta_0)\pk^{(-)}_\te(k), \\
	C^{EB}_\ell&=&\displaystyle\int dk \Delta^{E}_{\ell,\te}(k,\eta_0)\Delta^{B}_{\ell,\te}(k,\eta_0)\pk^{(-)}_\te(k),
\end{eqnarray}
with 
\begin{equation}
	\pk^{(\pm)}_\te(k)=\pk^R_\te(k)\pm\pk^L_\te(k).
\end{equation}
Following Ref. \cite{contaldi_etal_2008,magueijo_benincasa_2011,bethke_magueijo_2011a,bethke_magueijo_2011b}, the primordial power spectra of the left-handed and right-handed gravitational waves differ by two different effective Newton constants. As a consequence, one expect a change in amplitude but identical spectral indices for $\pk^R_\te$ and $\pk^L_\te$, {\it i.e.} $r_R\neq r_L$ and $n_R=n_L$. The same modifications are also obtained in the framework of pseudoscalar inflation \cite{sorbo_2011,anber_sorbo,cook_sorbo}. We subsequently model the primordial power spectra by
\begin{equation}
	\pk^{(\pm)}_\te(k)=r_{(\pm)}\times\mathcal{A}_\sa\times\left(\frac{k}{k_0}\right)^{n_\te},
\end{equation}
with $\mathcal{A}_\sa$ the amplitude of the power spectrum for scalar perturbations at the pivot scale, $k_0$, (set equal to $0.002$~Mpc$^{-1}$ in our study) and, $n_\te(=n_R=n_L)$ the tilt of the tensor modes. The parameters $r_{(\pm)}=r_R\pm r_L$ stand for the tensor-to-scalar ratio amounting the amplitude of $\pk^{(\pm)}_\te$. The parameter $r_{(+)}$ is positive-valued while $r_{(-)}$ can be either positive-valued ($r_R>r_L$) or negative-valued ($r_R<r_L$). Since the $BB$ correlations are only generated by $\pk_{(+)}$ and the $TB$ and $EB$ correlations by $\pk_{(-)}$, the amplitude of $C^{BB}_\ell$ measures the cosmological parameter $r_{(+)}$, while the amplitudes of $C^{TB}_\ell$ and $C^{EB}_\ell$ measure the parameter $r_{(-)}$. In a parity invariant universe, $r_R=r_L$ and one easily obtains $r_{(+)}= r$, the standard tensor-to-scalar ratio, and $r_{(-)}=0$. We stress that there is a priori no reason for $r_{(+)}$ to be equal to the tensor-to-scalar ratio of standard cosmology, $r$, except in the case of parity invariant universe. However, what is constrained thanks to a measurement of $C^{BB}_\ell$ is $r_{(+)}$ and from that perspective, $r_{(+)}$ plays the same role as $r$. 

Parity breaking is amounted by the parameter:
\begin{equation}
	\delta=\frac{r_{(-)}}{r_{(+)}}=\frac{r_{R}-r_{L}}{r_{R}+r_{L}},
\end{equation}
which varies from $-1\leq\delta\leq1$ since both $r_R$ and $r_L$ are greater than or equal to zero. Parity is not broken by gravity if $\delta=0$. The case of no production of left-handed(right-handed resp.) gravitational waves corresponds to $\delta=1(-1$ resp.). 

Moreover, the opposite convention of $\delta$ can be adopted, as in \cite{xia_2012}. It simply changes the sign of the $EB$ and $TB$ correlations. Indeed, through parity transformation, {\it i.e.}
\begin{eqnarray}
	r_{(+)}\to r'_{(+)}=r_{(+)}&\mathrm{and}&\delta\to\delta'=-\delta,
\end{eqnarray}
(corresponding to $r_R\to r'_R=r_L$ and $r_L\to r'_L=r_R$), the primary CMB anisotropies are changed to
\begin{eqnarray}
	C^{TB}_\ell&\to& {C'}^{TB}_\ell=-C^{TB}_\ell, \\
	C^{EB}_\ell&\to& {C'}^{EB}_\ell=-C^{EB}_\ell,
\end{eqnarray}
leaving the four other power spectra unchanged.

\subsection{Impact of lensing}

During their propagation from recombination to today, CMB photons travel through the potential-well of large scale structures deforming their trajectories because of gravitational lensing. This distorts the spatial distribution of primary anisotropies and deforms their angular power spectra. However, the gravitational lensing is usually neglected as mentioned in Ref.\cite{saito_2007}. We propose here to derive the impact of lensing by large scale structure and to show the obtained lensed power spectra. To this end, we adopt the harmonic formalism developed in Ref. \cite{hu_2000}, extended here to account for the presence of primary $TB$ and $EB$ correlations which are non-zero (see also Ref. \cite{challinor_lewis_2005,fabbian} for a real-space formalism). This computation is explicitly given in App. \ref{app:lensing} and we only provide here the final results. For temperature, one obtains :
 \begin{equation}
	\tilde{C}^{TT}_\ell=\left[1+R^{T}\right]C^{TT}_\ell+\displaystyle\sum_{\ell_1,\ell_2}F^{T}_{\ell\ell_1\ell_2}C^{\phi\phi}_{\ell_1}C^{TT}_{\ell_2}
\end{equation}
with
\begin{eqnarray}
	R^{T}&=&-\frac{1}{2}\ell(\ell+1)\displaystyle\sum_{\ell_3}\ell_3(\ell_3+1)\frac{2\ell_1+1}{4\pi}C^{\phi\phi}_{\ell_3}, \\
	F^{T}_{\ell\ell_1\ell_2}&=&\frac{1}{4}\left[\ell_1(\ell_1+1)+\ell_2(\ell_2+1)-\ell(\ell+1)\right]^2 \\
	&\times&{\frac{(2\ell_1+1)(2\ell_2+1)}{4\pi}}\left(\begin{array}{ccc}
			\ell & \ell_1 & \ell_2 \\
			0 & 0 &0
		\end{array}\right)^2. \nonumber
\end{eqnarray}

More interestingly is the case of the cross-correlation between temperature and polarization fields including primary $TB$ correlations :
\begin{eqnarray}
	\tilde{C}^{TE}_\ell&=&\left[1+R^{X}\right]C^{TE}_\ell+\displaystyle\sum_{\ell_1,\ell_2}F^{X}_{\ell\ell_1\ell_2}C^{\phi\phi}_{\ell_1}C^{TE}_{\ell_2}, \\
	\tilde{C}^{TB}_\ell&=&\left[1+R^{X}\right]C^{TB}_\ell+\displaystyle\sum_{\ell_1,\ell_2}F^{X}_{\ell\ell_1\ell_2}C^{\phi\phi}_{\ell_1}C^{TB}_{\ell_2},
\end{eqnarray}
with
\begin{eqnarray}
	F^{X}_{\ell\ell_1\ell_2}&=&\frac{1}{8}\left[\ell_1(\ell_1+1)+\ell_2(\ell_2+1)-\ell(\ell+1)\right]^2 \\
	&\times&{\frac{(2\ell_1+1)(2\ell_2+1)}{4\pi}}\left(\begin{array}{ccc}
			\ell & \ell_1 & \ell_2 \\
			0 & 0 &0
		\end{array}\right) \nonumber \\
	&\times&\left[\left(\begin{array}{ccc}
			\ell & \ell_1 & \ell_2 \\
			2 & 0 &-2
		\end{array}\right)\pm\left(\begin{array}{ccc}
			\ell & \ell_1 & \ell_2 \\
			-2 & 0 &2
		\end{array}\right)\right]. \nonumber
\end{eqnarray}
and
\begin{equation}
	R^{X}=-\frac{1}{2}\left[\ell(\ell+1)-2\right]\displaystyle\sum_{\ell_3}\ell_3(\ell_3+1)\frac{2\ell_3+1}{4\pi}C^{\phi\phi}_{\ell_3}.
\end{equation}
It is worth mentionning that the primary $TE$ angular power spectrum does {\it not} contribute to the lensed $TB$ angular power spectrum. If it were not the case, the former power spectrum would have spoilt the lensed $\tilde{C}^{TB}_\ell$ (at least in some range of angular scales). Indeed, Primary $TB$ correlations are only sourced by tensor modes in chiral gravity while primary $TE$ are sourced by both scalar and tensor modes. As a consequence, the polarized anisotropies are such as: $\left|C^{TE}_\ell\right|\gg \left|C^{TB}_\ell\right|$. This means that any leakages of primary $C^{TE}_\ell$ into $\tilde{C}^{TB}_\ell$ are a non-negligeable, if not dominant, contaminant of the primary $TB$ cross-correlation. As primary $TE$ are not affected by parity breaking, this would have significantly lowered the efficiency of using the (necessarily lensed) $TB$ angular power spectrum as a probe of chiral gravity.

Finally, the lensed angular power spectra for polarized anisotropies read
\begin{eqnarray}
	\tilde{C}^{EE}_\ell&=&\left[1+R^{P}\right]C^{EE}_\ell+\displaystyle\sum_{\ell_1,\ell_2}F^{(+)}_{\ell\ell_1\ell_2}C^{\phi\phi}_{\ell_1}C^{EE}_{\ell_2} \\
	&+&\displaystyle\sum_{\ell_1,\ell_2}F^{(-)}_{\ell\ell_1\ell_2}C^{\phi\phi}_{\ell_1}C^{BB}_{\ell_2} \nonumber \\
	\tilde{C}^{BB}_\ell&=&\left[1+R^{P}\right]C^{BB}_\ell+\displaystyle\sum_{\ell_1,\ell_2}F^{(+)}_{\ell\ell_1\ell_2}C^{\phi\phi}_{\ell_1}C^{BB}_{\ell_2} \\
	&+&\displaystyle\sum_{\ell_1,\ell_2}F^{(-)}_{\ell\ell_1\ell_2}C^{\phi\phi}_{\ell_1}C^{EE}_{\ell_2} \nonumber \\
	\tilde{C}^{EB}_\ell&=&\left[1+R^{P}\right]C^{EB}_\ell \\
	&+&\displaystyle\sum_{\ell_1,\ell_2}\left(F^{(+)}_{\ell\ell_1\ell_2}-F^{(-)}_{\ell\ell_1\ell_2}\right)C^{\phi\phi}_{\ell_1}C^{EB}_{\ell_2}, \nonumber
\end{eqnarray}
with
\begin{eqnarray}
	F^{(\pm)}_{\ell\ell_1\ell_2}&=&\frac{1}{16}\left[\ell_1(\ell_1+1)+\ell_2(\ell_2+1)-\ell(\ell+1)\right]^2  \nonumber \\
	&\times&{\frac{(2\ell_1+1)(2\ell_2+1)}{4\pi}} \\
	&\times&\left[\left(\begin{array}{ccc}
			\ell & \ell_1 & \ell_2 \\
			2 & 0 &-2
		\end{array}\right)\pm\left(\begin{array}{ccc}
			\ell & \ell_1 & \ell_2 \\
			-2 & 0 &2
		\end{array}\right)\right]^2. \nonumber
\end{eqnarray}
and
\begin{equation}
	R^P=-\frac{1}{2}\left[\ell(\ell+1)-4\right]\displaystyle\sum_{\ell_3}\ell_3(\ell_3+1)\frac{2\ell_3+1}{4\pi}C^{\phi\phi}_{\ell_3}.
\end{equation}
As is the case for $TB$ correlations, the lensed $EB$ power spectrum is not affected by the primary $EE$ and $BB$ power spectra. This once again means that the potential observation of a (necessarily lensed) non-vanishing $EB$ angular power spectrum is a direct tracer of non-zero $EB$ correlations prior to lensing. In the more precise setting of this study, this means that observing non-vanishing $EB$ (as well as non-vanishing $TB$) is a direct view of primary $EB$ cross-correlations due to parity breaking, though $\ell$-modes are reshuffled by lensing.

\subsection{Numerical results}
The explicit computation of the six angular power spectra is done by numerically solving for the Boltzmann equations. To this end, we modified the {\sc class} Boltzmann code \footnote{\tt http://class-code.net} \cite{class} incorporating two different primordial power spectra for left-handed and right-handed tensor modes, as well as the impact of lensing on primary anisotropies using the above-derived formulas. Our main interest is the $TB$ and $EB$ angular power spectra and we only show our results for such $C_\ell$'s (alongside the $BB$ spectrum used as a reference to evaluate the amplitude of the odd power spectra). The case of $TT$, $EE$, $BB$ and $TE$ power spectra is identical to standard, parity-invariant cosmology, setting $r=r_{(+)}$.
\begin{figure}
\begin{center}
	\includegraphics[scale=0.5]{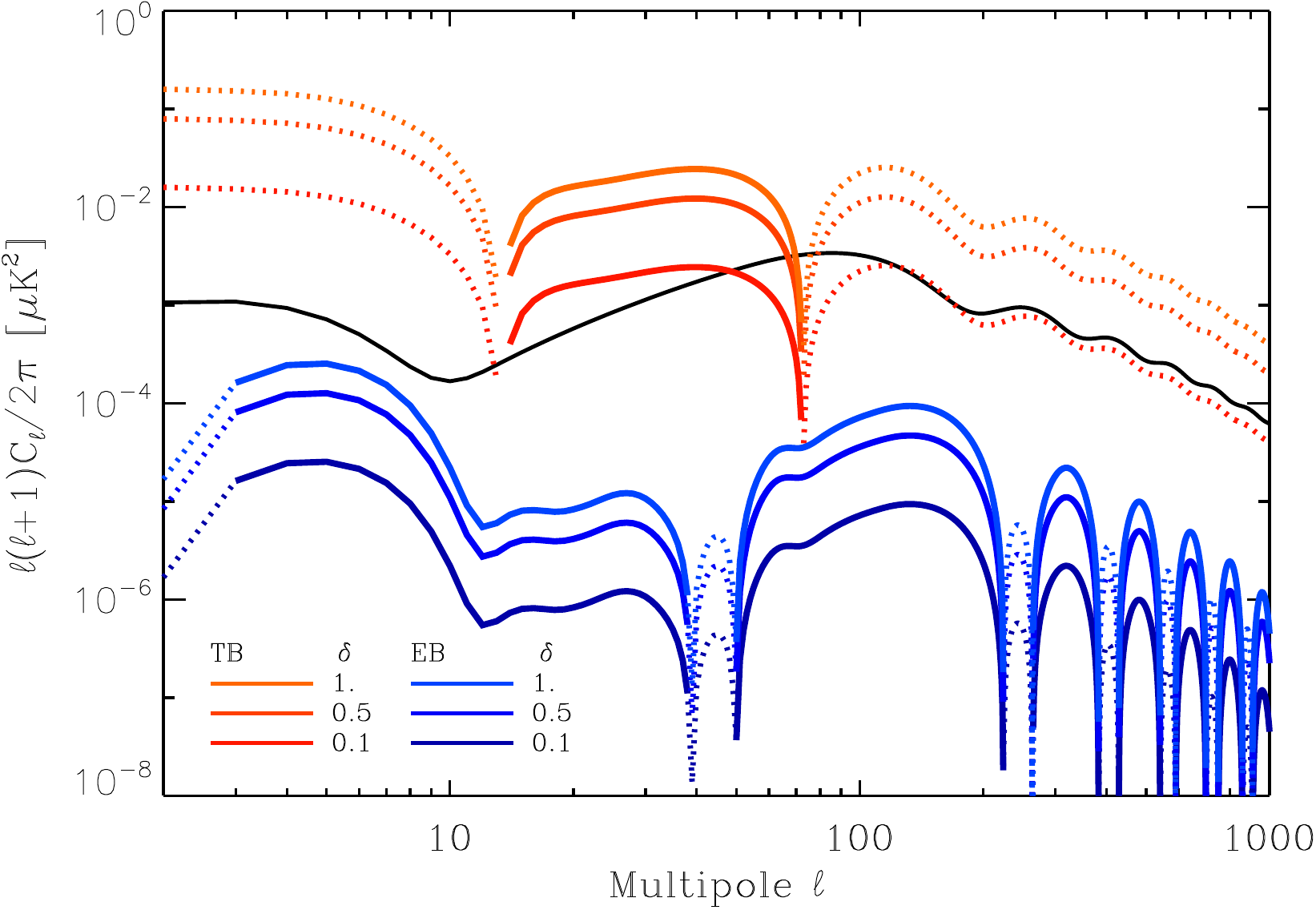}\\
	\includegraphics[scale=0.5]{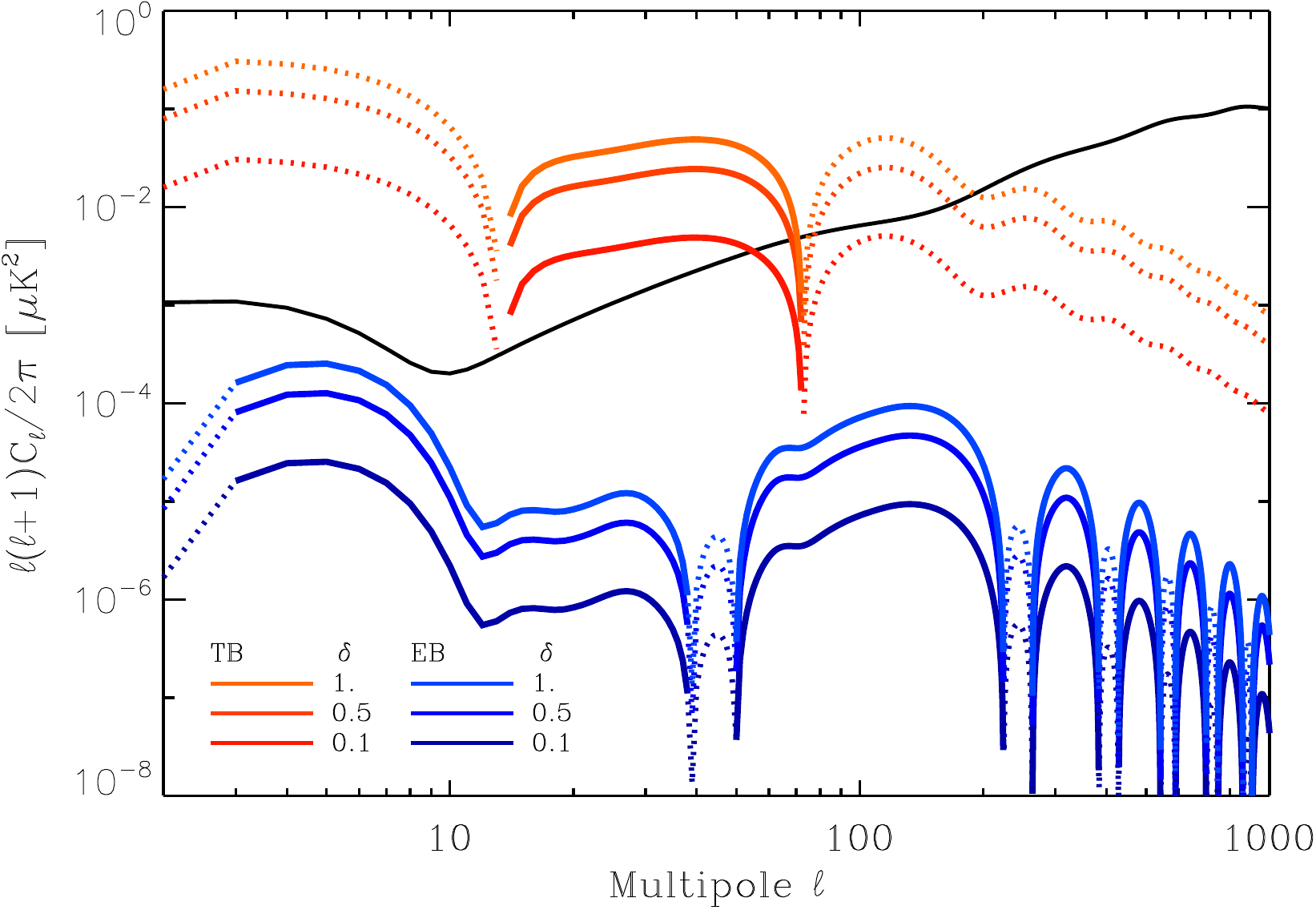}
	\caption{{\it Upper panel:} Angular power spectra for {\it primary} CMB anisotropies for $BB$ (black curve), $TB$ (red curves) and $EB$ (blue curves) correlations. The parameters $r_{(+)}$ is set equal to 0.05 and $\delta$ varies from 0.1 (meaning 10\% of parity violation) to 1 (100\% of parity violation). Solid lines correspond to positive values of the angular power spectra and dashed lines correspond to negative values. Changing from $(\delta)$ to $(-\delta)$ with $r_{(+)}$ unchanged changes the sign of $C^{TB}_\ell$ and $C^{EB}_\ell$ and leaves $C^{BB}_\ell$ unaffected. We note that smaller $\left|\delta\right|$ translates into smaller $\left|r_{(-)}\right|$. {\it Lower panel:} Same as upper panel but taking into account the impact of gravitational lensing.}
	\label{fig:primcl}
\end{center}
\end{figure}

The CMB angular power spectra for $BB$ (black curve), $TB$ (red-orange curves) and $EB$ (blue curves) in the case of {\it primary} anisotropies are depicted in the upper panel of Fig. \ref{fig:primcl}, for $n_R=n_L=0$. The parameter $r_{(+)}$ is set equal to 0.05 and $\delta=0.1,~0.5$ and 1, corresponding to $r_{(-)}=0.005,~0.025$ and 0.05. The specific case of $\delta=1$ corresponds to 100\% of parity violation. As $\delta$ is positive-valued, this parity break is in the right-handed sector, meaning that left-handed tensor modes are not produced at all. The shape of the power spectrum for negative values of $\delta$ is easily inferred from using the transformation rule of the $C_\ell$'s under parity transformation: changing from $(\delta)$ to $(-\delta)$ with $r_{(+)}$ unchanged changes the sign of $C^{TB}_\ell$ and $C^{EB}_\ell$ and leaves $C^{BB}_\ell$ unaffected. For $\delta>0$, the $TB$ angular power spectrum is negative at the largest scale, $\ell\leq10$, and the $EB$ spectrum is negative-valued for $\ell=2$. From the transformation rule under parity, this means that for negative values of $\delta$, the $TB$ angular power spectrum is {\it positive} for multipoles smaller than 10 while the $EB$ quadrupole becomes {\it positive}. The impact of lensing on the $TB$ and $EB$ power spectra is shown in the lower panel of Fig. \ref{fig:primcl}. As already underlined, gravitational lensing has only a mild impact on the odd-parity angular power spectra.

\section{Uncertainties on angular power spectrum reconstruction}
\label{sec:uncer}
\subsection{Experimental setups}
\label{sec:xp}
For numerical investigations, we define two fiducial experimental setups. Though idealized, they are chosen to reflect the general characteristics of forthcoming CMB experiments dedicated to $B$-modes detection. Those characteristics which crucially impact on the angular power spectrum reconstruction are the noise level, the beam width and a peculiar sky coverage.

\begin{figure}
\begin{center}
	\includegraphics[scale=0.3]{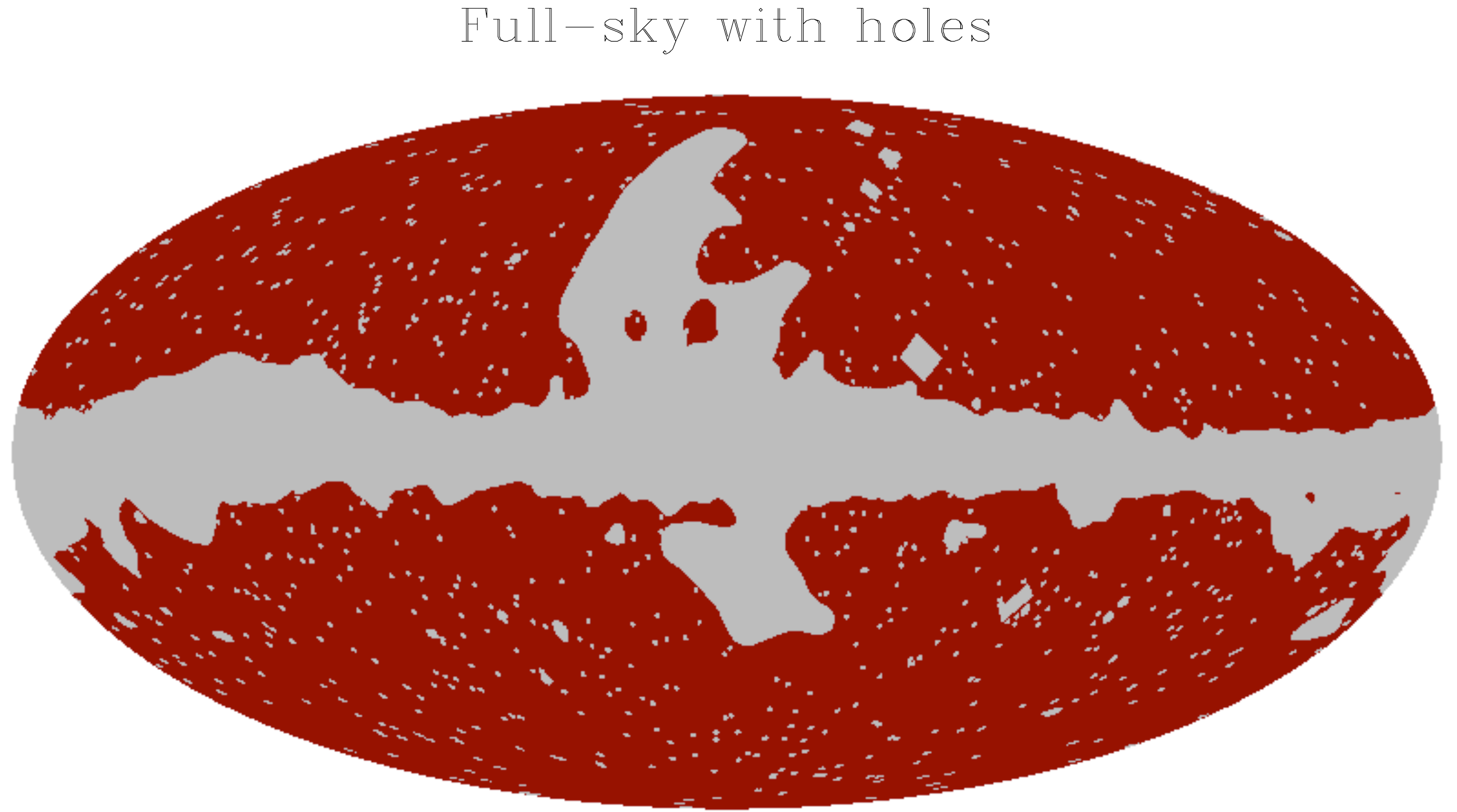} \\ \vspace*{0.3cm}
	\includegraphics[scale=0.3]{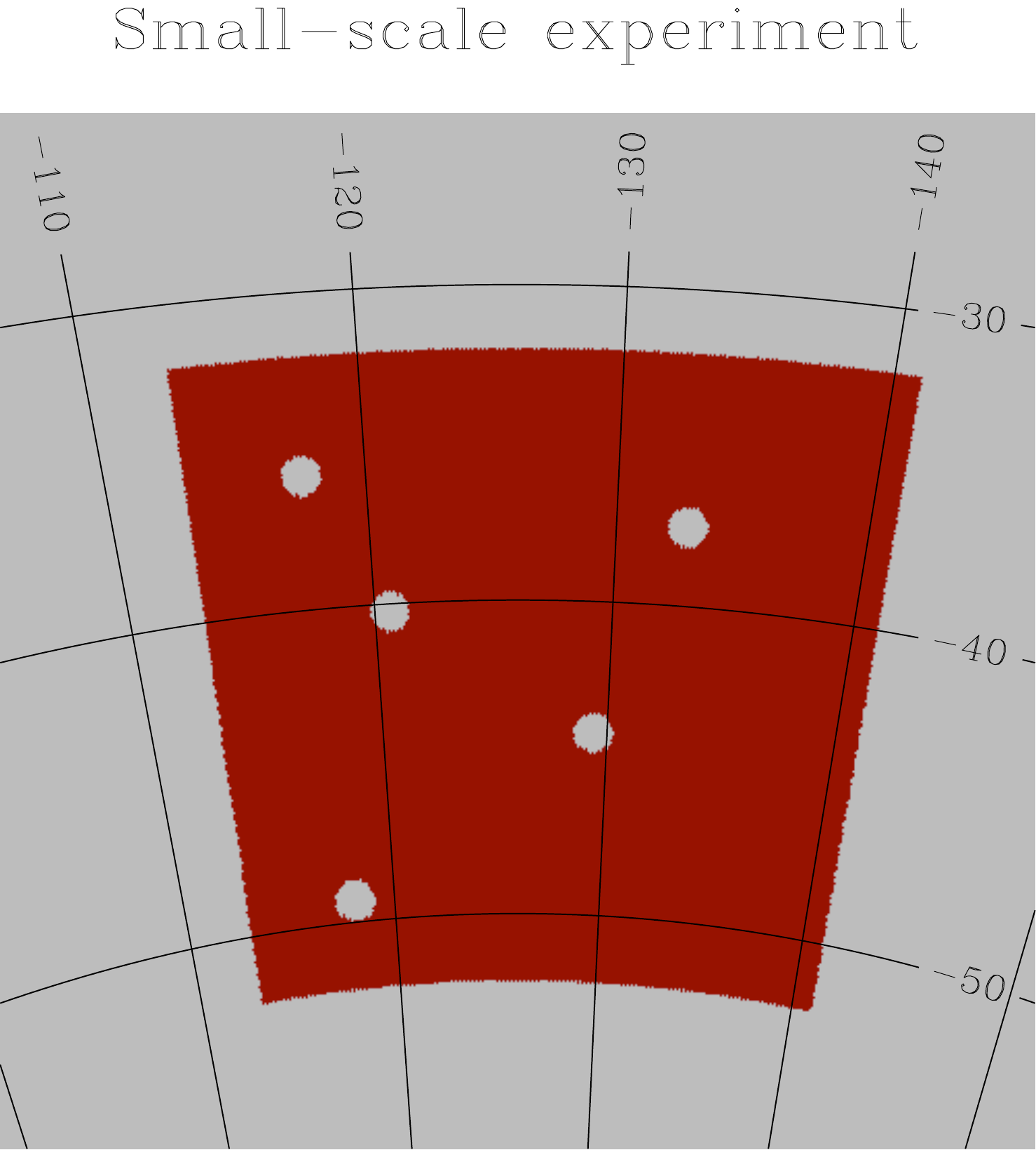}
	\caption{{\it Upper panel:} Sky area as observed by the fiducial satellite-like experiment as considered in this work. The sky coverage is $\sim71\%$ of the total celestial sphere. The mask is a combination of the galactic mask $R9$ and the point-sources catalog used for polarized data in \textsc{wmap} 7yr release. {\it Lower panel:} Sky area as observed by the fiducial balloon-borne, small-scale experiment as considered in this work. The sky coverages is $\sim1\%$ of the total celestial sphere.}
	\label{fig:masksat}
\end{center}
\end{figure}

First, we consider the case of a possible satellite experiment aimed at {\it primordial} $B$-mode detection. For such an experiment, we relied on the {\sc{epic}}$-2m$~\cite{bock_etal_2008} specifications for the noise level and the beam width, setting these to $2.2~\mu K$-arcmin for the noise level and $8~$arcmin for the beam width. For the peculiar sky coverage of such a 'nearly full-sky' experiment, we consider the galactic mask $R9$ used for polarized data in \textsc{wmap} 7yrs release (see \cite{wmap_7yr}) adding the point-sources catalog mask. So we obtain a $\sim71\%$ sky coverage patch showed in the upper panel of Fig.~\ref{fig:masksat}. 

Second, we consider the case of small-scale experiments inspired by the {\sc ebex}, ballon-borne experiment \cite{britt_etal_2010}. The noise level and the beam width are respectively set equal to $5.75~\mu K$-arcmin and $8~$arcmin. The observed part of the sky covers $\sim1\%$ of the total celestial sphere and its peculiar shape is displayed in the lower panel of Fig.~\ref{fig:masksat}. It consists of a square patch of an area of $\sim400$ square degrees including holes to mimic polarized point-sources removal. 

\subsection{Analytical and numerical error bars}
We use two approaches to derive the error bars and the covariance matrix on the estimated angular power spectra, denoted $\mathbf{\Sigma}$ in the following. The first one is based on a simple mode-counting. This underestimates the uncertainties as leakages due to cut-sky effects (and the full complexity of the mask) are not taken into account. Second, we make use of Monte-Carlo simulations using a {\it pure pseudospectrum} code for reconstructing the angular power spectra from the maps of the Stokes parameters \cite{grain_etal_2012}. 

\subsubsection{Mode-counting expressions}
The na\"\i{v}e mode counting derivation of the covariance matrix leads to :
\begin{eqnarray}
	\left[\mathbf{\Sigma}\right]^{XY,X'Y'}_{\ell,\ell'}&=&\left<C^{XX'}_\ell C^{YY'}_{\ell'}\right>-\left<C^{XX'}_\ell\right>\left<C^{YY'}_{\ell'}\right> \label{eq:cov1} \\
	&=&\delta_{\ell,\ell'}\left(\frac{1}{(2\ell+1)f_\mathrm{sky}}\right) \nonumber \\
	&\times&\left[\left(C^{XX'}_\ell+\frac{N^{XX'}_\ell}{B^2_\ell}\right)\left(C^{YY'}_\ell+\frac{N^{YY'}_\ell}{B^2_\ell}\right)\right. \nonumber \\
	&+&\left.\left(C^{XY'}_\ell+\frac{N^{XY'}_\ell}{B^2_\ell}\right)\left(C^{YX'}_\ell+\frac{N^{YX'}_\ell}{B^2_\ell}\right)\right], \nonumber
\end{eqnarray}
with $X,~X'$ and $Y,~Y'$ taking the values $T,~E$ and $B$. In the above formulas, the quantities $f_{\text{sky}}$, $B_\ell$ and $N^{XX'}_\ell$ described the impact of the instrumental strategy on the CMB anisotropies reconstruction: $f_{\text{sky}}$ stands for the observed (or kept in the analysis) fraction of the sky, $B_\ell$ is the multipolar decomposition of the beam of the telescope, and, $N^{XX'}_\ell$ describes the instrumental noise power spectrum entering in the estimation of $\tilde{C}^{XY}_\ell$. As explained in the appendix C of Ref. \cite{grain_etal_2012}, this noise power spectrum vanishes for $TE$, $TB$ and $EB$ cross-correlations as long as the noise in the $I,~Q,$ and $U$ maps is not correlated between two different Stokes parameters. We will assume here that this is indeed the case.

It is worth to mention that because none of the angular power spectra are vanishing in this setting, the six angular power spectra shows cross-correlations. As an example, the correlations between the $BB$ and $TB$ estimators are given by :
\begin{equation}
	\left[\mathbf{\Sigma}\right]^{BB,TB}_{\ell\ell'}=\frac{2\delta_{\ell,\ell'}}{(2\ell+1)f_{\text{sky}}}\tilde{C}^{TB}_{\ell}\left[\tilde{C}^{BB}_\ell+\frac{N^{BB}_\ell}{B^2_\ell}\right],
\end{equation}
which is non zero for non-vanishing $C^{TB}_\ell$.

\subsubsection{Error bars from pure pseudospectrum}
We also estimate the statistical uncertainties on the reconstructed $C_\ell$'s using a more elaborated approach based on Monte-Carlo simulations. Though the above-mentioned formulas are easy to handle with, they underestimate the error bars expected using realistic statistical tools to reconstruct the angular power spectra from maps of the CMB sky. In particular, it neglects the impact of leakages due to cut-sky effects in the case of pseudospectrum-based estimation of the $C_\ell$'s. Those leakages between multipoles and between polarization modes (see {\it e.g.} \cite{bunn_2002}) increase the sampling variance of angular power spectra estimations. For the more specific case of $B$-modes, such an increase could be dramatic as the much higher $E$-mode leaks into the much weaker $B$-mode. Those leakages can be corrected on average \cite{hauser_peebles_1973,hinshaw_etal_2003} but, if not corrected at the level of the variances of the estimators, the much higher $C^{EE}_\ell$ and $C^{TE}_\ell$ will inevitably contribute to the sampling variance of $C^{BB}_\ell$, $C^{TB}_\ell$ and $C^{EB}_\ell$, thus significantly increasing it. We therefore rely on the {\sc x}$^2${\sc pure} code \cite{grain_etal_2012} which implements the so-called {\it pure} pseudospectrum estimators, correcting for $E/B$ mixing on average {\it and} at the level of variances \cite{smith_2006}. The pure decomposition of the polarization field introduced in Ref. \cite{bunn_etal_2003} allows $E$ and $B$-modes to be exactly separated on a partial sky and for any single realization of the CMB polarized anisotropies, a sufficient condition for removing any $E$-modes leaking into $B$ for all the statistical moments of the angular power spectra estimators. A prescription based on the pure decomposition has been introduced in Ref. \cite{smith_2006} (and later elaborated on in Refs. \cite{smith_zaldarriaga_2007,grain_etal_2009}) to built a pseudospectrum estimator for $C^{BB}_\ell$ free of any $E/B$ mixing. This approach has finally been extended to incorporate odd power spectra in Ref. \cite{grain_etal_2012} (see also Ref. \cite{louis_etal_2013} for a flat sky implementation of the pure pseudospectrum estimator). 

It was shown that the so-called hybrid computation\footnote{The so-called hybrid computation means that angular power spectra are estimated using {\it pure} pseudomultipoles of $B$-types and {\it standard} pseudomultipoles of $E$-type.}  is the most accurate for estimating $BB$, $TB$ and $EB$ angular power spectra for the case of small-scale experiments covering $\sim1\%$ of the celestial sphere, and assuming a parity invariant universe, {\it i.e.} $C^{TB}_\ell=C^{EB}_\ell=0$ \cite{grain_etal_2012}. More recently, the need for such a pure approach in the context of satellite experiments, allowing for an estimation of the $C_\ell$'s over $\sim71\%$ of the celestial sphere, has been proved for the specific case of the $BB$ angular power spectrum \cite{ferte_etal_2013}. Such a pseudospectrum approach is therefore a method of a choice for analyzing forthcoming data currently taken by small-scale experiments as well as a potential, long-term, satellite mission dedicated to $B$-mode. 

We use the {\sc x}$^2${\sc pure} code in a Monte-Carlo setting to derive realistic estimates of the statistical uncertainties (including sampling variance and noise variance) for the two above-defined different experimental configurations. The angular power spectra are estimated within bandpower with the first bin ranging from $\ell=2$ to $\ell=20$ and the following bins having a width of $\Delta_b=40$.

\subsection{Results for the two experimental configurations}
\begin{figure*}
\begin{center}
\begin{tabular}{cc}
	\includegraphics[scale=0.4]{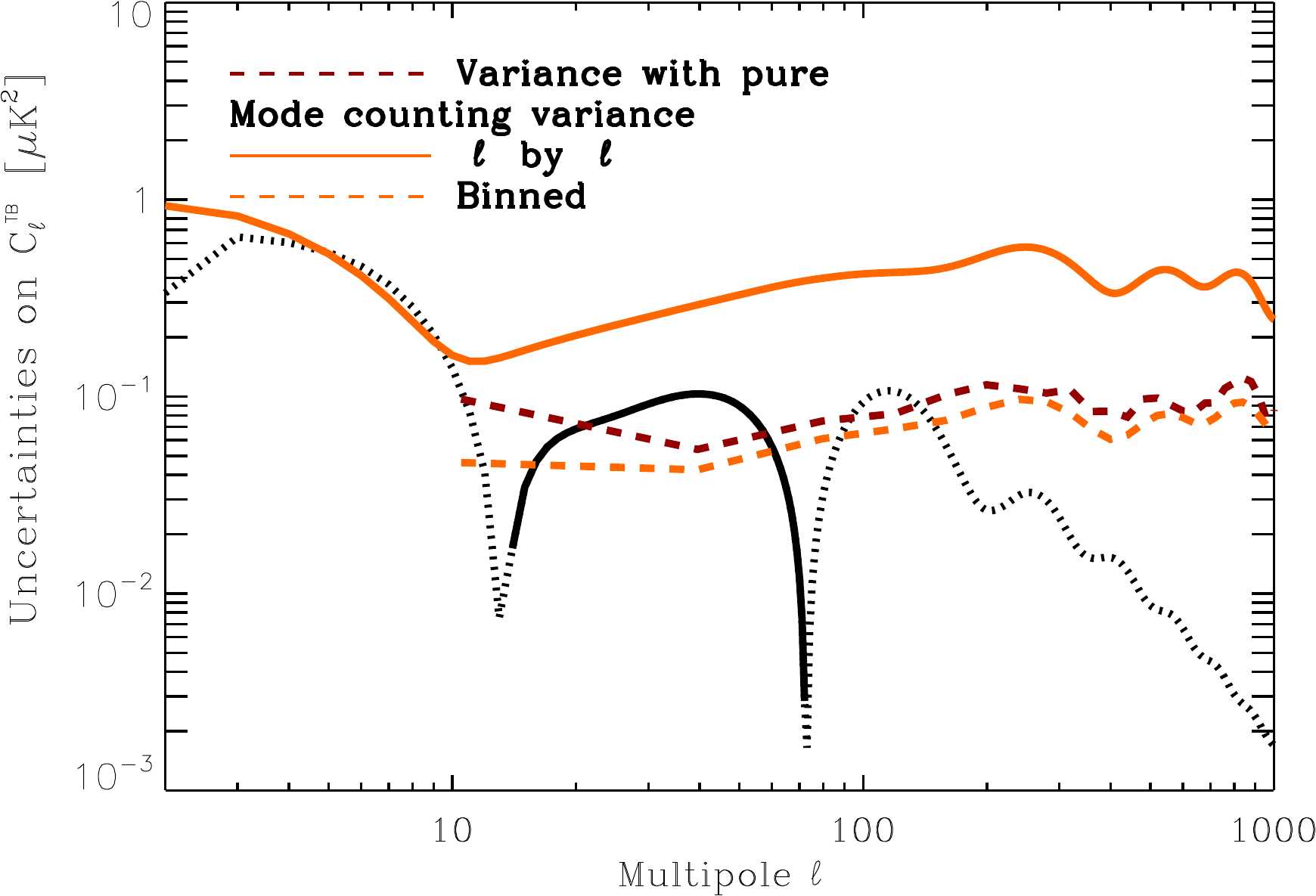}&\includegraphics[scale=0.4]{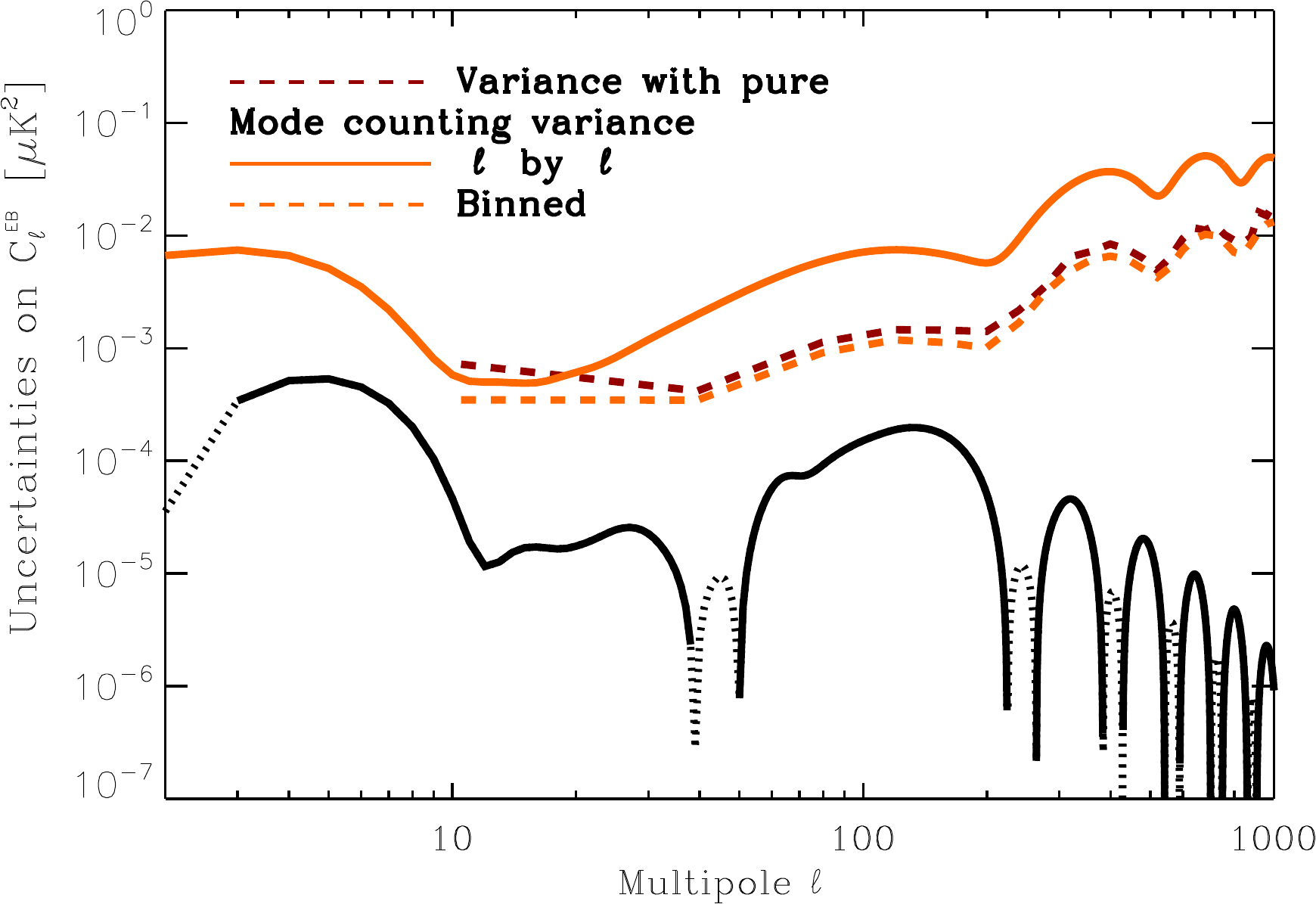} \\
	\includegraphics[scale=0.4]{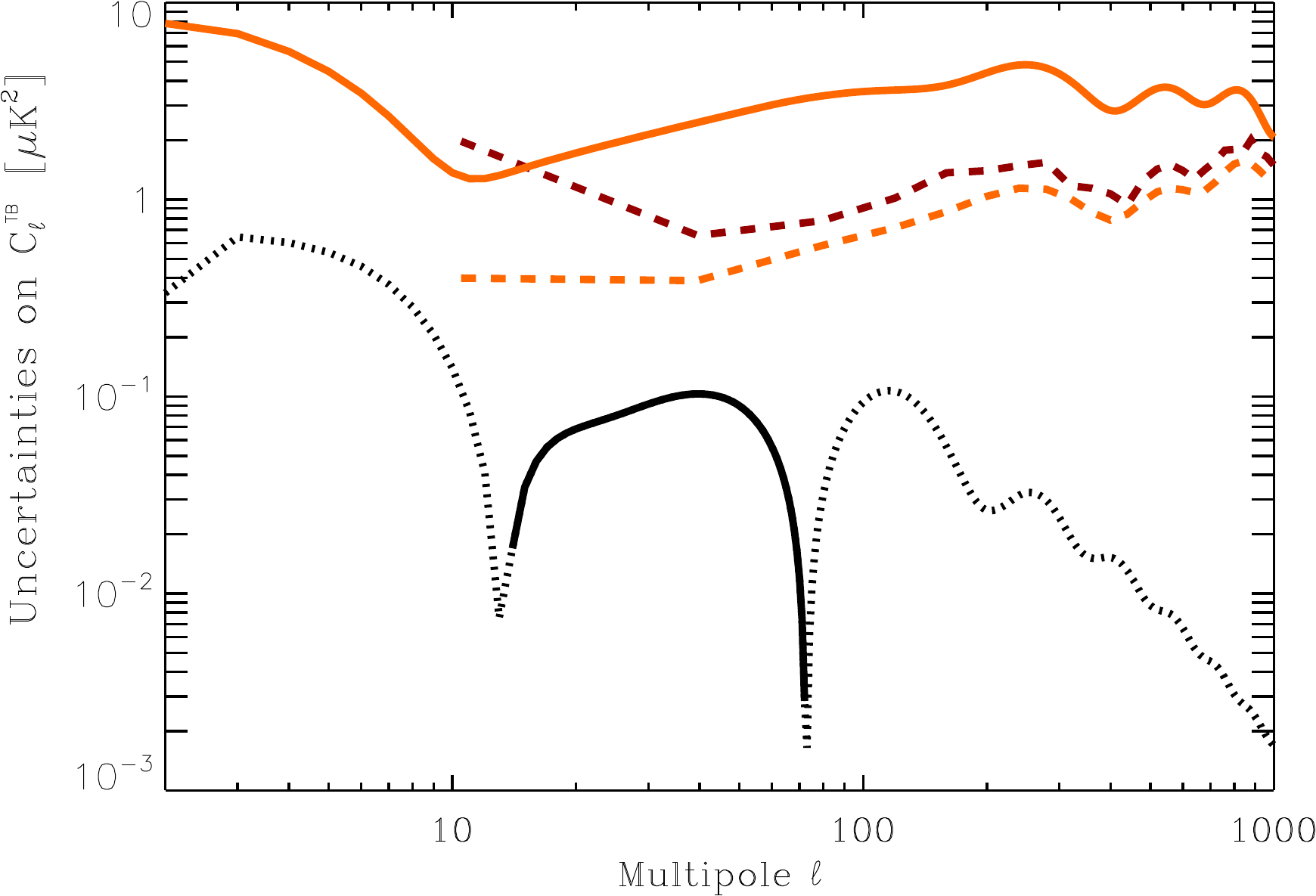}&\includegraphics[scale=0.4]{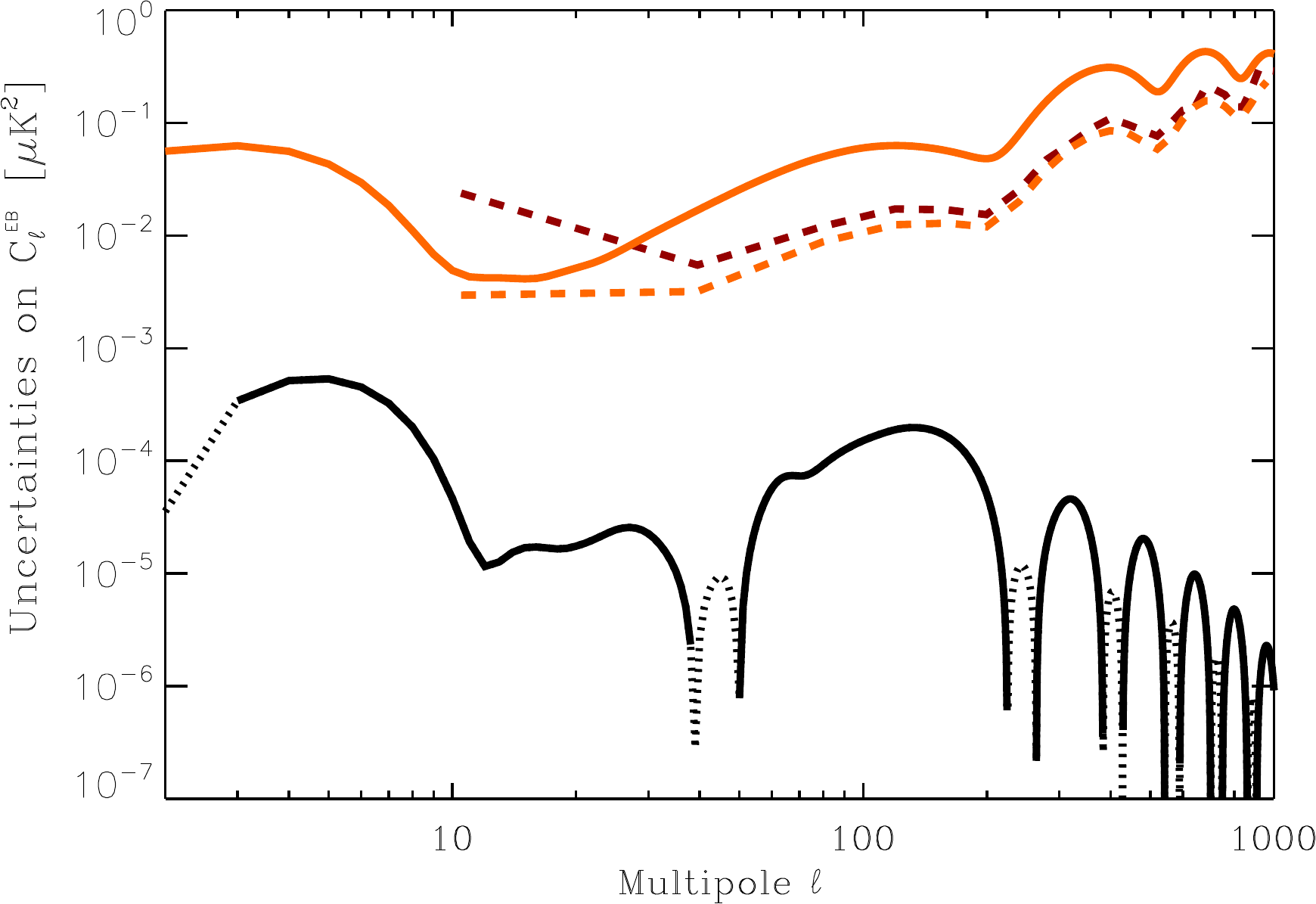}
\end{tabular}
	\caption{Uncertainties on the reconstructed $C^{TB}_\ell$ (left panels) and $C^{EB}_\ell$ (right panels) angular power spectra for two experimental configurations: a satellite mission covering $\sim71\%$ of the sky (upper panels) and a small-scale experiment covering $\sim1\%$ of the sky (lower panels). Black curves are the input angular power spectra. Solid-orange curves stands for the error bars using an $\ell$-by-$\ell$ mode counting derivation. Dashed-red curves are the error bars obtained from 500 Monte-Carlo simulations using the pure pseudospectrum reconstruction of the angular power spectra. This has to be compared to the {\it binned} mode-counting computation of those error bars given by the dashed-orange curves.}
	\label{fig:errorpur}
\end{center}
\end{figure*}

Our numerical results for the uncertainties on the reconstruction of the odd-power spectra are depicted in Fig. \ref{fig:errorpur} for the two above-described experimental configurations, and the two approaches to compute the error bars. Black curves are the input angular power spectra. Solid-orange curves stand for the error bars using an $\ell$-by-$\ell$ mode counting derivation, Eq. (\ref{eq:cov1}). Dashed-red curves are the error bars obtained from 500 Monte-Carlo simulations using the pure pseudospectrum reconstruction of the angular power spectra. This has to be compared to the {\it binned} mode-counting computation of those error bars given by the dashed-orange curves, {\it i.e.}
\begin{equation}
	\left[\mathbf{\Sigma}\right]^{A,A}_{b,b}=\ds\sum_{\ell\in b}\left[\frac{\ell(\ell+1)}{2\pi\Delta_b}\right]^2\left[\mathbf{\Sigma}\right]^{A,A}_{\ell\ell},
\end{equation}
where we used the fact that the mode-counting covariance is diagonal in the $\ell$-space. 

We consider here the case of $r_{(+)}=0.1$ and $\delta=1$ ({\it i.e.} $r_{(-)}=0.1$), in line with the latest constraints on the tensor-to-scalar ratio \cite{planck_inf,biceppaper}.

\subsubsection{Analytical error bars}
As shown in Sec. \ref{sec:cell}, the $TB$ angular power spectra is higher in amplitude than the $BB$ spectrum, and one could be tempted to deduce that detecting $TB$ cross-correlations would be easier than detecting $BB$ correlations. However, one can expect rather high error bars on the reconstructed $\tilde{C}^{TB(EB)}_\ell$ at large angular scales simply because of the sampling variance. Indeed the sampling variance for the $TB$ correlations, reads
\begin{equation}
	\left[\mathbf{\Sigma}\right]^{TB,TB}_{\ell\ell}=\frac{1}{\left(2\ell+1\right)f_\text{sky}}\left[\left(\tilde{C}^{TB}_\ell\right)^2+\tilde{C}^{TT}_\ell\tilde{C}^{BB}_\ell\right]. \nonumber
\end{equation}
The $TB$ and $BB$ spectra are sourced by tensor perturbations {\it only}. However, the $TT$ spectrum is generated by {\it both} scalar and tensor perturbations. We therefore expect $\sqrt{\tilde{C}^{TT}_\ell\tilde{C}^{BB}_\ell}\gg\tilde{C}^{TB}_\ell$ and the signal-to-noise ratio for $TB$ roughly scales as
\begin{equation}
	\frac{\tilde{C}^{TB}_\ell}{\sqrt{\left[\mathbf{\Sigma}\right]^{TB,TB}_{\ell\ell}}}\sim\sqrt{\left(2\ell+1\right)f_\text{sky}}\left(\frac{\tilde{C}^{TB}_\ell}{\sqrt{\tilde{C}^{TT}_\ell\tilde{C}^{BB}_\ell}}\right) \nonumber
\end{equation}
which is much smaller than unity because $C^{TT}_\ell$ is sourced by scalar perturbations\footnote{For $BB$, the signal-to-noise ratio is given by $\sqrt{(\ell+1/2)f_\text{sky}}$ for such an ideal case dominated by the sampling variance. Clearly, detecting $\tilde{C}^{BB}_\ell$ is easier than detecting $\tilde{C}^{TB}_\ell$ though the later is higher in amplitude than the former.}. The same argument applies to the case of the $EB$ cross-correlations as scalar perturbations contribute via $\tilde{C}^{EE}_\ell$. 

This is clearly highlighted in Fig.\ref{fig:errorpur}, focusing on the orange curve. For the noise levels considered here, the uncertainties are completely dominated by the sampling variance from $\ell=2$ to $\ell=1000$ and, more precisely, by the term $\sqrt{\tilde{C}^{TT(EE)}_\ell\tilde{C}^{BB}_\ell}$ for $\tilde{C}^{TB(EB)}_\ell$. Detecting the $TB$ and $EB$ angular power spectra {\it multipole by multipole } is impossible even in this rather optimistic case ($\delta=1$ and $r_{(+)}=0.1$) and one should rely on binning for a positive detection of such $C_\ell$'s in this framework.

\subsubsection{Error bars from pure pseudospectrum}
The error bars on the $C^{TB(EB)}_\ell$'s using the pure pseudospectrum estimation is depicted by the dashed-red curves. For the case of the satellite-like mission, the $TB$ angular power spectrum can be detected in the three first bins while the detection of $EB$ is not possible without a drastic increase of the width of the bandpowers. For the case of small-scale experiments, neither the $TB$ spectrum nor the $EB$ one can be measured, at least with the size of the bandpower here-adopted.

We note here that the expected uncertainties shown in Fig. \ref{fig:errorpur} are very high so that one could be tempted to conclude that a detection of $r_{(-)}$ is impossible. Nevertheless, the fact that detecting $\tilde{C}^{TB(EB)}_\ell$ multipole by multipole (or bin by bin) is not possible {\it does not necessarily} imply that detecting $r_{(-)}$ is impossible, for the detection of such a parameter is done by resumming the angular power spectrum over bandpowers. Since those bandpowers are assumed to be uncorrelated, this will inevitably decrease the uncertainty on $r_{(-)}$ by a factor $\sim\sqrt{N}$, with $N$ the total number of resummed bins.

\section{Forecasts on chiral gravity}
\label{sec:detect}
\subsection{Fisher matrix formalism}
\subsubsection{Fisher matrix}
Detecting chiral gravity using CMB polarized anisotropies translates into the possible detection of non-vanishing CMB angular power spectra of odd type, and subsequently into the possible measurement of $r_{(-)}$ from those odd power spectra. To this end, we will rely on a simple Fisher analysis \cite{fisher} to translate the uncertainties on the odd power spectrum reconstruction into errors on the recovery of $r_{(-)}$ (see also Ref. \cite{stivoli} for a more elaborated approach). Such an approach has already been proved to be useful in such a context for {\it e.g.} forecasting constraints on bouncing cosmology in loop quantum cosmology \cite{grain_etal_2010}. 

As explained in Sec. \ref{sec:uncer}, the six {\it estimated} CMB angular power spectra are cross-correlated and this additional source of information should be kept in the Fisher analysis. We therefore use the six angular power spectra as our "input"~data and define the Fisher information matrix as follows :
\begin{equation}
	\left[\mathbf{F}\right]_{ij}=\ds\sum_{A,A'}\sum_{b,b'}\left.\frac{\partial C^A_b}{\partial\theta_i}\right|_{\bar{\theta}_i}\times\left[\mathbf{\Sigma}^{-1}\right]^{A,A'}_{b,b'}\times\left.\frac{\partial C^{A'}_{b'}}{\partial\theta_j}\right|_{\bar{\theta}_j},
	\label{eq:fishermat}
\end{equation}
where the $A$ and $A'$ superscripts runs over $TT,~EE,~BB,~TE,~TB$ and $EB$, and $b,~b'$ denote the bandpowers. Our set of parameters is $\theta_i\equiv(r_{(+)},r_{(-)})$ and the above Fisher information matrix is just the inverse of the covariance matrix for $\theta_i$ assuming the likelihood to be gaussian. The {\it marginalized} signal-to-noise ratio $\mathrm{(S/N)}_{\theta_i}$ for a given parameter $\theta_i$ is finally given by $\mathrm{(S/N)}_{\theta_i}=\bar\theta_i/\sqrt{[\mathbf{F}^{-1}]_{ii}}$.

It is worth mentioning that though only $C^{TB}_\ell$ and $C^{EB}_\ell$ do depend on $r_{(-)}$, the constraints that can be set on that parameter using the {\it six} angular power spectra will be different than the constraint obtained by using {\it solely} the two odd-parity angular power spectra. In the latter case, the covariance matrix entering the Fisher information matrix would be a sub-block of the full covariance matrix. However, the inverse of that sub-block is not equal to the sub-block of the inverse of the full covariance as long as the estimated $TB$ and $EB$ power spectra are correlated to the other spectra, which is indeed the case here. 

\subsubsection{Parameterizing the input angular power spectra}
Information about $r_{(+)}$ is in principle contained in the six angular power spectra. However, the $TT,~EE$ and $TE$ correlations are mainly generated by the scalar inhomogeneities and we can safely set those angular power spectrum to their best-fit shape and considered them as independent of $r_{(+)}$. This approximation is valid as in this study, we will consider values of $r_{(+)}$ smaller than 0.2. Information about $r_{(-)}$ is solely contained in the $TB$ and $EB$ cross-correlations. We therefore modeled the CMB anisotropies as $\tilde{C}^{BB}_\ell=f(r_{(+)})$ as a function of $r_{(+)}$, and, $\tilde{C}^{TB(EB)}_\ell=f(r_{(-)})$ as functions of $r_{(-)}$. The parameters $r_{(\pm)}$ provide the global amplitude of the $B$-related angular power spectra. Those power spectra can therefore be parametrized as an amplitude, given by $r_{(+)}$ for $\tilde{C}^{BB}_\ell$ and given by $r_{(-)}$ for $\tilde{C}^{TB(EB)}_\ell$, multiplied by a template (plus a constant term coming from lensing for the specific case of the $BB$ spectrum). The different templates will be denoted using calligraphic font, $\mathcal{T}$.

As a function of $r_{(+)}$, the lensed $BB$ angular power spectrum reads
\begin{equation}
	\tilde{C}^{BB}_\ell[r_{(+)}]=r_{(+)}\times \mathcal{T}^{BB}_{\ell}+ \mathcal{T}^{EE\to BB}_{\ell,lens},
\end{equation}
with $\mathcal{T}^{BB}_{\ell}$ and $\mathcal{T}^{EE\to BB}_{\ell,lens}$ two fiducial angular power spectra, independent of $r_{(+)}$, and given by
\begin{eqnarray}
	\mathcal{T}^{BB}_{\ell}&=&\left[1+R^{P}\right]C^{BB}_\ell[r_{(+)}=1] \\
	&+&\displaystyle\sum_{\ell_1,\ell_2}F^{(+)}_{\ell\ell_1\ell_2}C^{\phi\phi}_{\ell_1}C^{BB}_{\ell_2}[r_{(+)}=1], \nonumber
\end{eqnarray}
and
\begin{equation}
	\mathcal{T}^{EE\to BB}_{\ell,lens}=\displaystyle\sum_{\ell_1,\ell_2}F^{(-)}_{\ell\ell_1\ell_2}C^{\phi\phi}_{\ell_1}C^{EE}_{\ell_2}.
\end{equation}
The fiducial $C^{BB}_\ell[r_{(+)}=1]$ is easily computed using the line of sight solution and setting $r_{(+)}=1$. The $EE$ angular power spectra involved in $\mathcal{C}^{BB}_{\ell,lens}$ is obtained using the Planck best fit.

The two odd angular power spectra are similarly expressed using two fiducial power spectra which do not depend on $r_{(-)}$, {\it i.e.}
\begin{equation}
	\tilde{C}^{TB(EB)}_\ell[r_{(-)}]=r_{(-)}\times \mathcal{T}^{TB(EB)}_{\ell},
\end{equation}
with
\begin{eqnarray}
	\mathcal{T}^{TB}_{\ell}&=&\left[1+R^{TB}\right]C^{TB}_\ell[r_{(-)}=1] \\
	&+&\displaystyle\sum_{\ell_1,\ell_2}F^{TB}_{\ell\ell_1\ell_2}C^{\phi\phi}_{\ell_1}C^{TB}_{\ell_2}[r_{(-)}=1], \nonumber
\end{eqnarray}
and
\begin{eqnarray}
	\mathcal{T}^{EB}_{\ell}&=&\left[1+R^{P}\right]C^{EB}_\ell[r_{(-)}=1] \\
	&+&\displaystyle\sum_{\ell_1,\ell_2}\left(F^{(+)}_{\ell\ell_1\ell_2}-F^{(-)}_{\ell\ell_1\ell_2}\right)C^{\phi\phi}_{\ell_1}C^{EB}_{\ell_2}[r_{(-)}=1]. \nonumber
\end{eqnarray}
As is the case for the $BB$ angular power spectrum, the two fiducial power spectra $C^{TB(EB)}_\ell[r_{(-)}=1]$ are easily computed using the line of sight solution and setting $r_{(-)}$ equal to 1.

The parameter $r_{(+)}$ plays the same role as $r$ in standard cosmology. As a consequence, any observational constraint on $r$ can be directly translated into a constraint on $r_{(+)}$. We remind that from temperature as measured by the  {\sc planck} satellite, the tensor-to-scalar ratio is bounded from above: $r<0.11$ at 95\% CL \cite{planck_inf}, while the latest measurement of polarization by the {\sc bicep2} experiment constrains $r=0.2^{+0.07}_{-0.05}$ \cite{biceppaper}. This latest constraint on the tensor-to-scalar ratio probably needs some confirmation. As a consequence, we will explore values of $r_{(+)}$ ranging from 0.007 to 0.2 with a specific focus on the $r_{(+)}=0.05,~0.1$ and $0.2$

\subsection{Detection of $r_{(-)}$ : satellite mission}
\label{sec:satres}
\subsubsection{Results with mode-counting covariance}
\label{sssec:modecount}
We first consider the case of the signal-to-noise ratio on the parameter $r_{(-)}$. A preliminary study is to inquire the values of the signal-to-noise ratio obtained on the above-mentioned parameters relying on a simple mode-counting error bars estimation. This warrants an efficient exploration of the measurable range of $r_{(-)}$ and $\delta$ using the {\sc x}$^2${\sc pure} code for a correct estimation of the uncertainties.

Let us first briefly mention that taking into account the $TT,~EE,~BB$ and $TE$ power spectra brings an additional amount of information thus increasing the signal-to-noise ratio on $r_{(-)}$. This additional piece of informations simply consists in the fact that those angular power spectra do {\it not} depend on $r_{(-)}$ and it is finally transferred into the ending values of $[\mathbf{F}]_{r_{(-)},r_{(-)}}$ since the $TB$ and $EB$ spectra are {\it correlated} to the four other ones. Without such correlations, adding $TT,~EE,~BB$ and $TE$ would have not change the signal-to-noise ratio. Considering the case of $r_{(+)}=r_{(-)}=0.1$, the derived signal-to-noise ratio on $r_{(-)}$ using solely $TB$ and $EB$ would be $\sim5$, to be compared to 5.6 using the full set of angular power spectra (see Tab. \ref{tab:snrrmlarge}).

The table \ref{tab:snrrmlarge} summarizes our results on $\mathrm{(S/N)}_{r_{(-)}}$ for different values of $r_{(+)}$ and $r_{(-)}$ (keeping in mind that $r_{(-)} \leq r_{(+)}$) using the mode-counting expressions for the covariance of the $C_\ell$ and marginalized over $r_{(+)}$. We note that the correlations between $r_{(-)}$ and $r_{(+)}$ are small for the ranges of values here-explored, {\it i.e.}
$$
\frac{\mathbf{F}_{r_{(+)},r_{(-)}}}{\sqrt{\mathbf{F}_{r_{(+)},r_{(+)}}~\mathbf{F}_{r_{(-)},r_{(-)}}}}\sim10^{-3}. 
$$
However, the signal-to-noise ratio for a fixed value of $r_{(-)}$ decreases for higher values of $r_{(+)}$ since the higher $r_{(+)}$, the higher $\tilde{C}^{BB}_\ell$ and the higher the uncertainties on $\tilde{C}^{TB(EB)}_\ell$. This translates inevitably into higher uncertainties on $r_{(-)}$. 

\begin{table}
\begin{tabular}{cc||c|c|c|c|c|c}
	& $r_{+}$ & $0.2$ & $0.1$ & $0.07$ &  $0.05$ & $ 0.03$ & $0.007$  \\
	$r_{-}$ & & & & & & & \\
	\hline\hline
	$0.2$ &&	\underline{10.6}	&				&					 &			       &               &           \\
	\hline
	$0.1$ && \underline{3.8} 	        & \underline{5.66}	& 			                 & 			       & 		&           \\
	\hline
	$0.07$&&	\underline{2.57}       &  \underline{3.6}  	& \underline{$4.3$}           &			       & 		&           \\
	\hline
	$0.05$&& 1.8			        &  \underline{2.46}	& \underline{$2.91$}	        & \underline{$3.4$} &   	         &           	\\
	\hline
	$0.03$&& 1.07 			        &  1.44 		  	& 1.68	    	                 & 1.95 	               & {\underline{2.44}}         &  	       \\
	\hline
	$0.007$&& 0.25                  	& 0.33              		& 0.39                               & 0.44                     & 0.54        &  0.94  \\
	\hline
\end{tabular}
	\caption{Signal-to-noise on $r_{(-)}$ for different values of $r_{(+)}$ in the case of satellite mission and using a mode-counting expression for the error bars on the angular power spectra reconstruction. The underlined values correspond to a detection at $2\sigma$ of parity violation.}
	\label{tab:snrrmlarge}
\end{table}

A $2\sigma$ detection of parity violation is guaranteed if $r_{(+)}\geq0.05$ {\it and} $r_{(-)}\geq0.05$. The values of $r_{(-)}=0.03$ appears as a threshold value since for $r_{(-)}<0.03$ the signal-to-noise ratio is systematically below unity. For this precise value of $r_{(-)}=0.03$, the signal-to-noise ratio on $r_{(-)}$ varies from 1.44 to 2.44, for $r_{(+)}=0.1$ (a parity violation of 30\%) and $r_{(+)}=0.03$ (a parity violation of 100\%), respectively.

Moreover, for a tensor-to-scalar ratio $r = 0.2$ as favored by the {\sc bicep2} experiment \cite{bicep2_2014}, a detection of chiral gravity with at least $2\sigma$ is expected for parity violation greater or equal to $35\%$.

\begin{figure}
\begin{center}
	\includegraphics[scale=0.5]{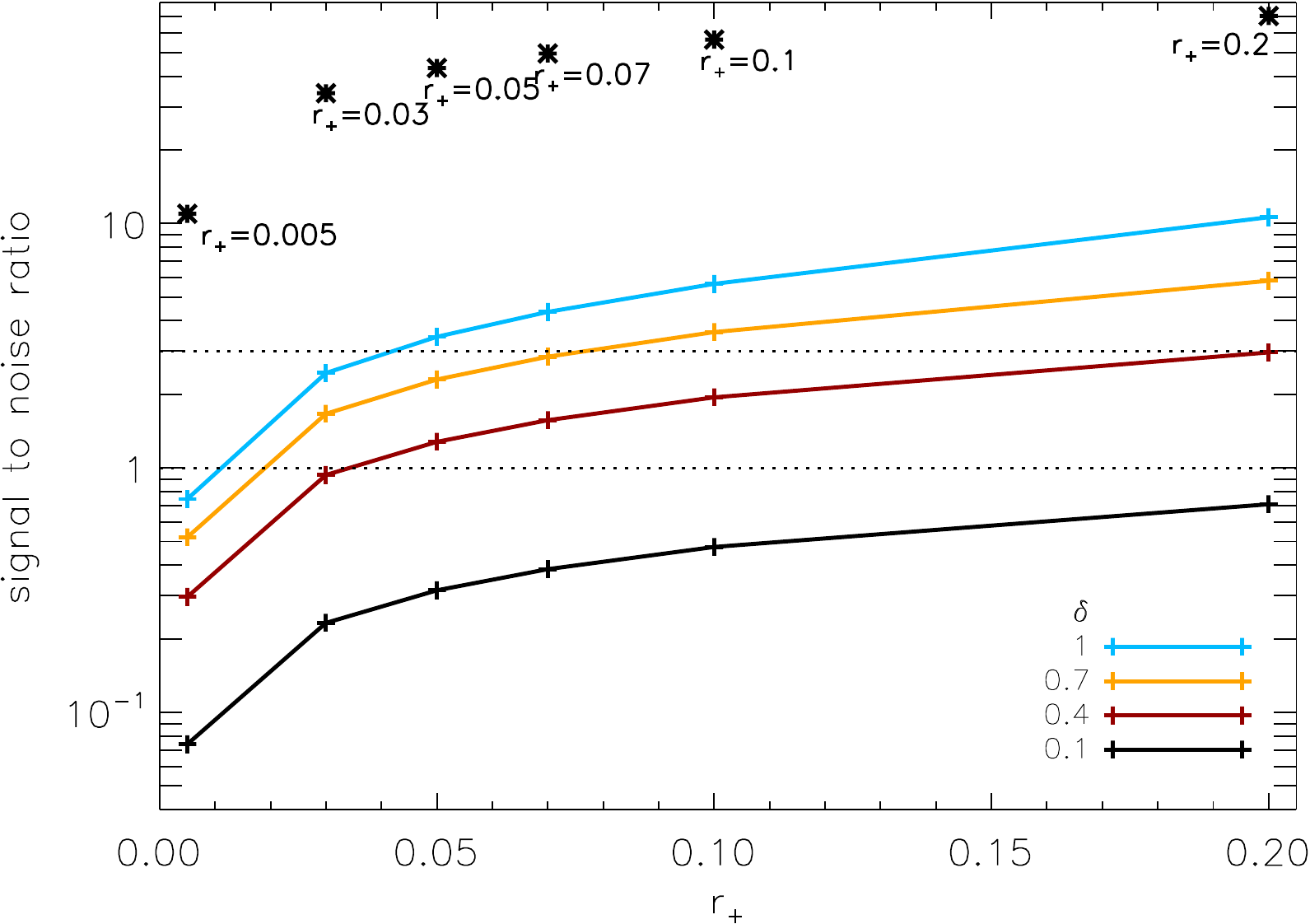} 
	\caption{Signal-to-noise ratio on the $r_+$ (black cross) and $r_-$ (colored lines) parameters is here depicted for four values of $\delta$. The black dashed lines figure the 1$\sigma$ and 3$\sigma$ level of detection.}
	\label{fig:snrsat}
\end{center}
\end{figure}

The figure \ref{fig:snrsat} shows the $\mathrm{(S/N)}_{r_{(+)}}$ (black crosses) and $\mathrm{(S/N)}_{r_{(-)}}$ (colored lines) as a function of $r_{(+)}$ and for different level of parity violation, $\delta$ ranging from 10\% to 100\%. (We remind that for a fixed value of $\delta$, the value of $r_{(-)}$ increases for higher values of $r_{(+)}$.) For $r_{(+)}\leq0.11$ (as favored by the {\sc planck} results on temperature anisotropies), a $3\sigma$ detection of parity violation can be achieved for $\delta\geq70\%$ and a $2\sigma$ detection is possible for a parity violation greater than 50\%, and a minimal value of $r_{(+)}\sim0.05$ appears as mandatory for such a detection. For $r_{(+)}=0.2$ (as favored by polarization measurements of {\sc bicep2}), $\delta=40\%$ could be detected at $3\sigma$.

\subsubsection{Results with pure pseudo-$C_\ell$ covariance}
\label{sssec:pseudo}
Based on this optimistic exploration of the detectable range of parity violation, we then estimate {\it realistic} statistical error bars on the reconstructed angular power spectra in the context of the pure pseudospectrum estimators. Our results are summarized in the table~\ref{tab:snr}, considering $r_{(+)}=0.05,~0.1$ and $0.2$ and $\delta=0.5$ and $1$.

For the most optimistic case, \textit{i.e.} $r_{(+)} = 0.2$ and $\delta = 1$, the obtained signal-to-noise ratio on the $r_{(-)}$ parameter is (S/N)$_{r_{(-)}} = 5.46$ then a detection at $5\sigma$ would be possible. This has to be compared to a 10$\sigma$ detection assuming the (underestimated) mode-counting derivation of the statistical uncertainties. If for that same value of $r_{(+)}=0.2$, parity violation is reduced to $\delta=0.5$ (corresponding to $r_{(-)}=0.1$), its detection is reduced by more than a factor 2, the signal-to-noise ratio on $r_{(-)}$ being $\sim$2.5. The same conclusions are drawn for the case of $r_{+} = 0.1$, the signal-to-noise ratio ranging from 3.67 for $\delta=1$ down to 1.51 for $\delta=0.5$, using the pure reconstruction of $B$-modes. We finally look at the case of $\delta = 1$ and $r_{(+)} = 0.05$. The obtained result is: (S/N)$_{r_{(-)}} = 2.35$ meaning a detection of chiral gravity of at least $2\sigma$ for $\delta=1$. (For the same situation and assuming the mode-counting estimation of the error bars, a 3$\sigma$ detection would have been inferred.) Similarly, if the level of parity violation is only of 50\% (corresponding to $r_{(-)}=0.025$), the signal-to-noise ratio is reduced by a factor $\sim2$.

We also consider the extreme case where parity is not violated, $\delta = 0$, and setting $r_{(+)} = 0.05$ and $r_{(+)}=0.2$. In the first case, the computed value of the uncertainty on the value of $r_{(-)}$ is $\sigma_{r_{(-)}} = 0.023$. This fixes a minimal detectable value of $r_{(-)} \sim 0.046$ at 95\% CL for such a possible satellite mission dedicated to $B$-mode, assuming $r_{(+)}=0.05$. In that case, detecting $EB$ and $TB$ power spectra compatible with zero corresponds to an upper bound on the level of parity violation of $\delta\leq0.92$ at 95\% CL. For $r_{(+)}=0.2$, this upper bound becomes $\delta\leq0.39$ at 95\% CL.
\begin{table}
\begin{center}
\begin{tabular}{cc||ccc|ccc}
	& & & & & & & \\
	& & & $\delta=1$ & & & $\delta=0.5$ & \\ \hline\hline
	& $r_{(+)}=0.2$ & & 5.46 & & & 2.5 & \\ \hline
	& $r_{(+)}=0.1$ & & 3.67 & & & 1.51 & \\ \hline
	& $r_{(+)}=0.05$ & & 2.35 & & & 1.11 & \\
	\hline
\end{tabular}
	\caption{Signal-to-noise ratio on $r_{(-)}$, (S/N)$_{r_{(-)}}$, as derived from a pure pseudospectrum reconstruction of the angular power spectra. We remind that for a given value of $r_{(+)}$ and $\delta$, the value of $r_{(-)}$ is $r_{(-)}=\delta\times r_{(+)}$.}
	\label{tab:snr}
\end{center}
\end{table}

\subsubsection{Impact of miscalibration angle}
There are many systematic effects affecting the reconstruction of the Stokes parameter starting from the time stream data. Among them, a miscalibration of the projection on the sky of the polarization orientation of the detectors will turn out into a rotation of the Stokes parameter, $P_\pm\to P^{\mathrm{(obs)}}_\pm=e^{\pm2i\Delta\psi}\times P_\pm$ \cite{hu_etal_2003,yadav_etal_2010,shimon_etal_2008}.  A way to estimate $\Delta\psi$ is to put the detecting $TB$ and $EB$ correlation equal to zero as they are expected to vanish in standard cosmology \cite{keating_etal_2013}. In the context of cosmological parity-violation parametrized by {\it e.g.} $r_{(-)}$, the miscalibration angle has to be estimated from $TB$ and $EB$ jointly to $r_{(-)}$. We propose to quantify how the miscalibration of the polarization angle can deteriorate the previous obtained constraints on chiral gravity. 
\begin{figure*}
\begin{center}
	\includegraphics[scale=0.35]{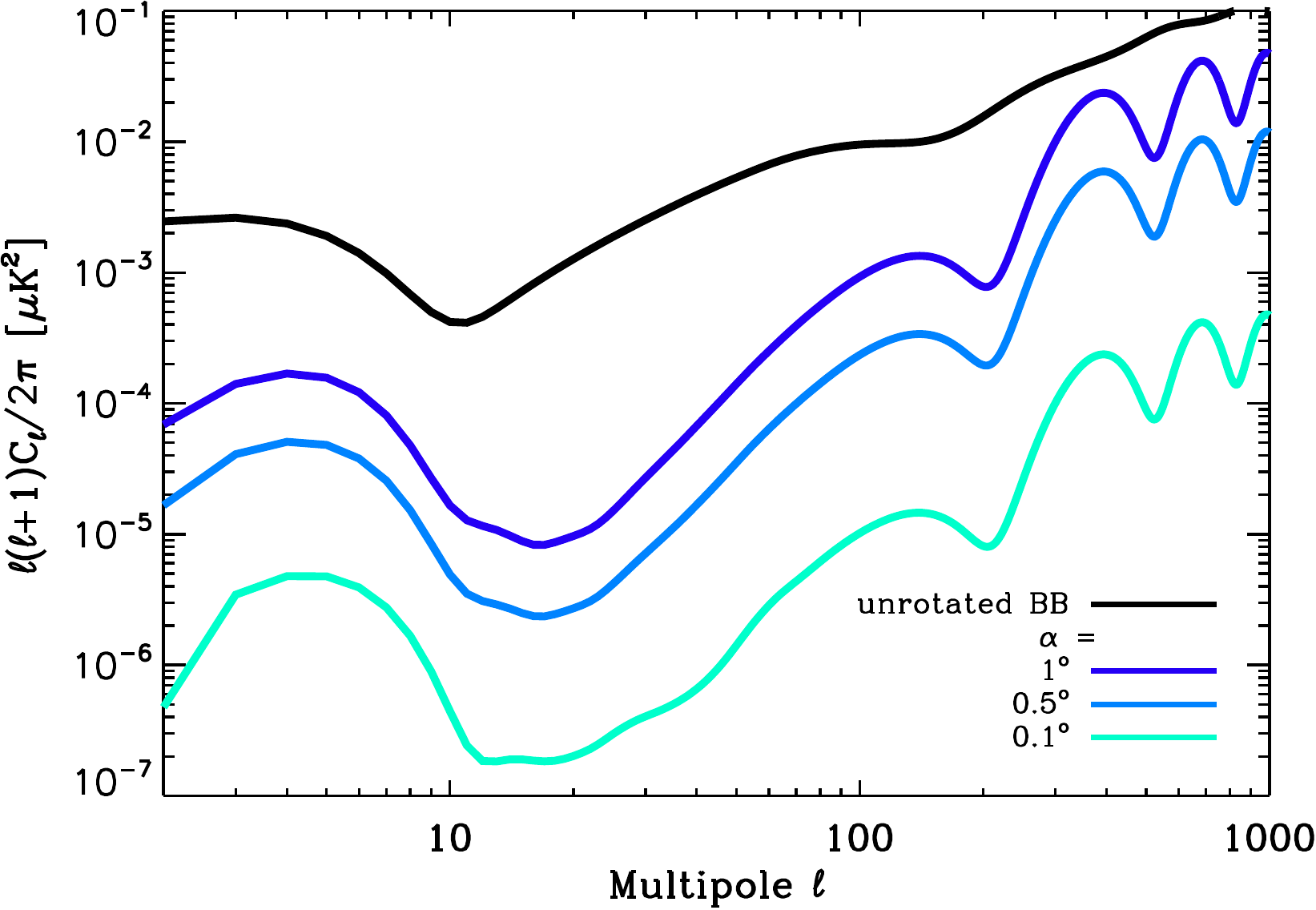} \includegraphics[scale=0.35]{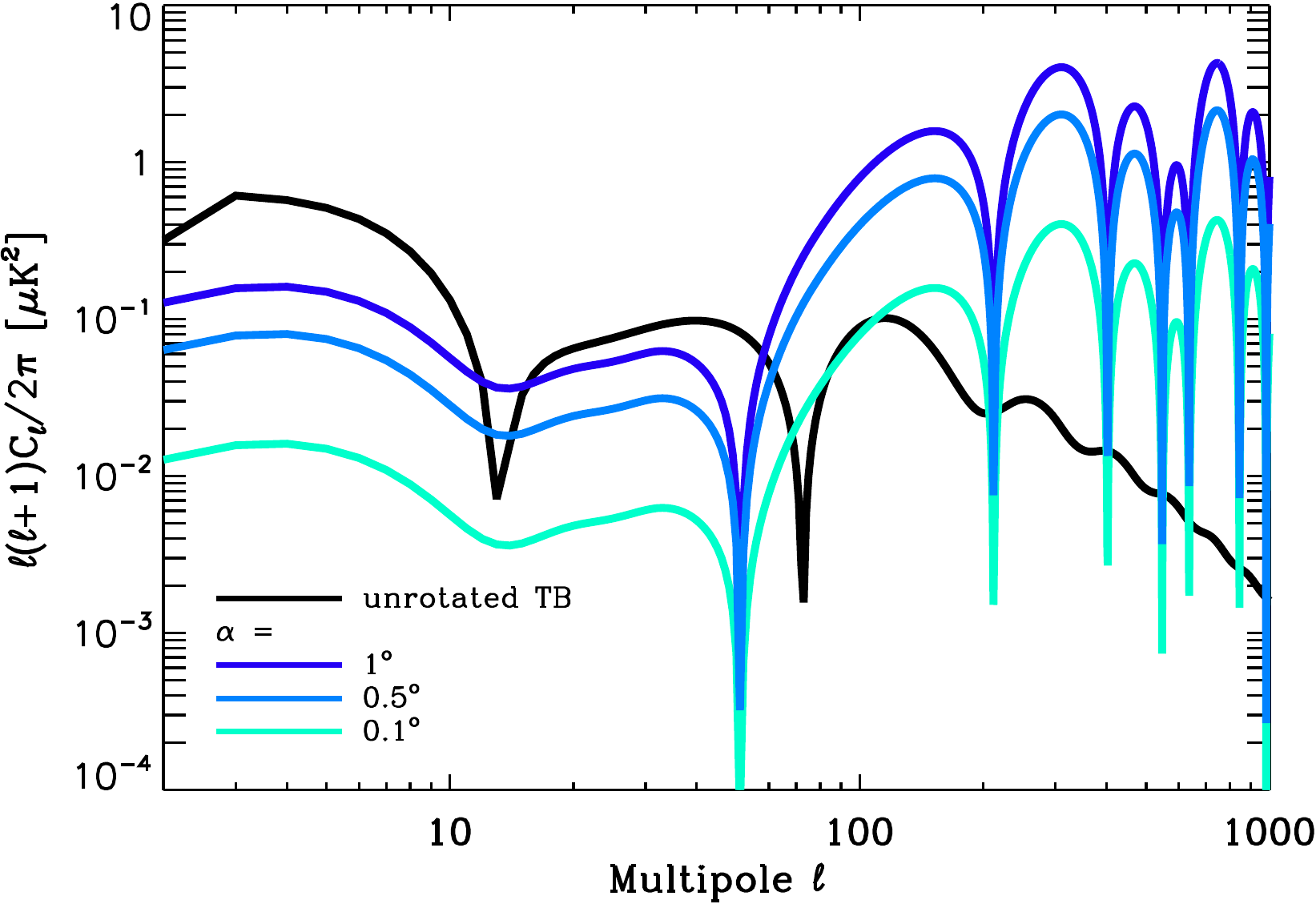} \includegraphics[scale=0.35]{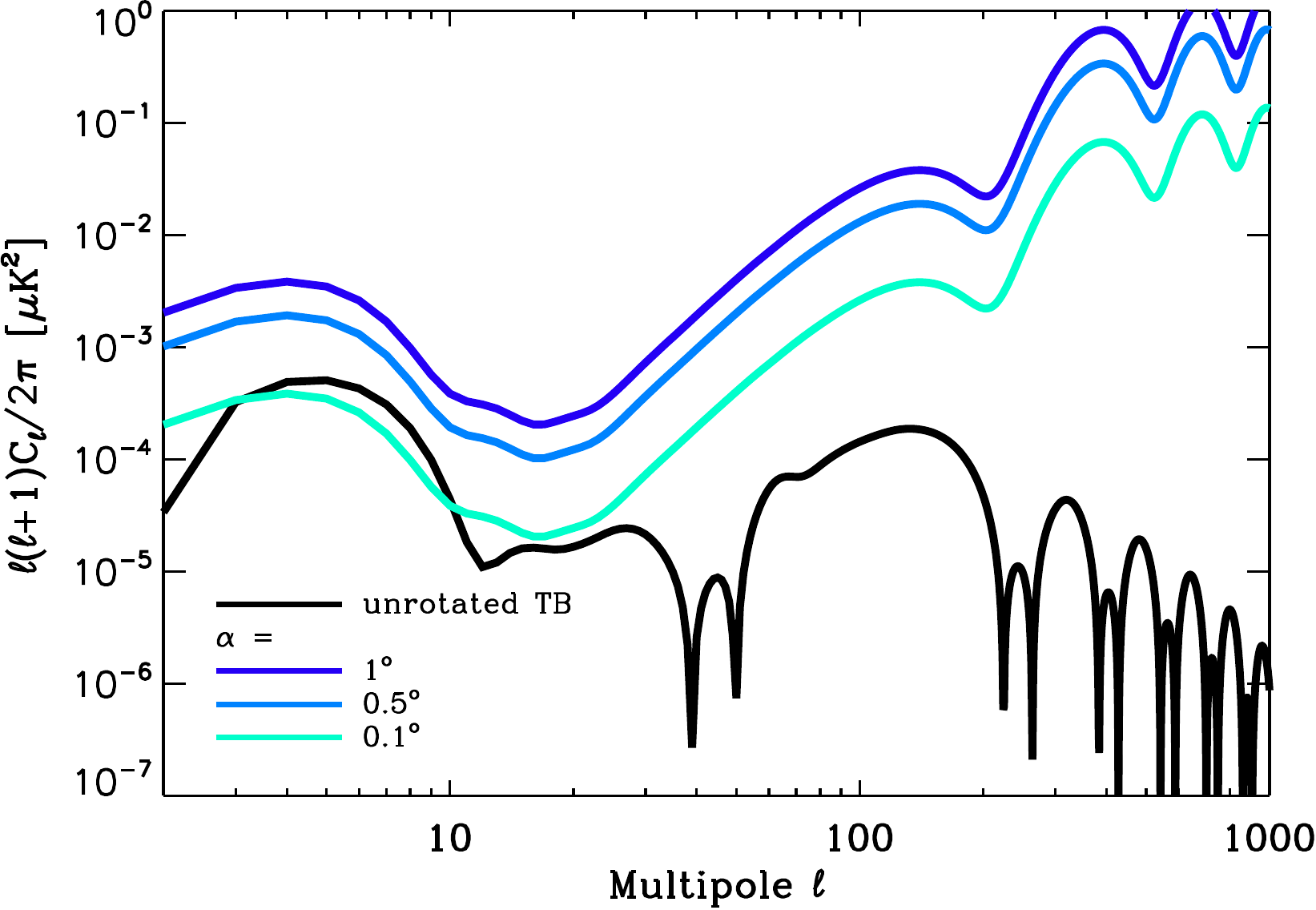}
	\caption{A miscalibration of the polarization angle impacts the amplitude and shape of the power spectra. The unrotated power spectra of the $B$-modes and the $TB$ and $EB$ cross-correlations are depict in black solid curve from left to right panels. The difference between the rotated power spectra and the unrotated is represented in dashed lines for different values of the miscalibration angle $\Delta\Psi$.}
	\label{fig:clrot}
\end{center}
\end{figure*}

As a consequence of this systematic effect, the observed angular power spectra are a linear combination of the real CMB angular power spectra. With non-vanishing $TB$ and $EB$ cross-correlations, the observed angular power spectra are :
\begin{eqnarray}
	\tilde{C}^{TT~\mathrm{(rot)}}_\ell&=&\tilde{C}^{TT}_\ell \\
	\tilde{C}^{TE~\mathrm{(rot)}}_\ell&=&\cos(2\Delta\psi)\tilde{C}^{TE}_\ell-\sin(2\Delta\psi)\tilde{C}^{TB}_\ell, \\
	\tilde{C}^{TB~\mathrm{(rot)}}_\ell&=&\sin(2\Delta\psi)\tilde{C}^{TE}_\ell+\cos(2\Delta\psi)\tilde{C}^{TB}_\ell, \\
	\tilde{C}^{EE~\mathrm{(rot)}}_\ell&=&\cos^2(2\Delta\psi)\tilde{C}^{EE}_\ell+\sin^2(2\Delta\psi)\tilde{C}^{BB}_\ell \\
	&+&\sin(4\Delta\psi)\tilde{C}^{EB}_\ell, \nonumber \\
	\tilde{C}^{BB~\mathrm{(rot)}}_\ell&=&\sin^2(2\Delta\psi)\tilde{C}^{EE}_\ell+\cos^2(2\Delta\psi)\tilde{C}^{BB}_\ell \\
	&-&\sin(4\Delta\psi)\tilde{C}^{EB}_\ell, \nonumber \\
	\tilde{C}^{EB~\mathrm{(rot)}}_\ell&=&\frac{1}{2}\sin(4\Delta\psi)\left(\tilde{C}^{EE}_\ell-\tilde{C}^{BB}_\ell\right) \\
	&+&\cos(4\Delta\psi)\tilde{C}^{EB}_\ell. \nonumber
\end{eqnarray}
As compared to the results shown in {\it e.g.} Ref. \cite{keating_etal_2013}, one can notice the additionnal contribution of $\tilde{C}^{TB}_\ell$ and $\tilde{C}^{EB}_\ell$. However, even in the case of vanishing $TB$ and $EB$ spectra, such miscalibration leads to spurious non-vanishing odd-parity angular power spectra. In Fig. \ref{fig:clrot}, the intrinsic CMB angular power spectra (solid-balck curves), $\tilde{C}_\ell$, and the leaked power due to rotation (dashed-colored curves), $\Delta C_\ell=(\tilde{C}^{\mathrm{(rot)}}_\ell-\tilde{C}_\ell)$, are displayed for the $BB$, $TB$ and $EB$ correlations and $\Delta\psi=0.1,~0.5$ and $1$ degree.

We note that the above modeling of the impact of miscalibrating the orientation of the detectors implicitly assumes that such angle is identical over the entire observed patch. If some variations are allowed, the resulting angular power spectra would be a convolution of the intrinsic CMB spectra with the angular power spectra of the rotation angle. This would significantly increase the complexity of the problem as this convolution mixes different multipoles. We also remind that the impact of homogeneous cosmic birefringence is exactly identical to the impact of a miscalibration of the polarizers orientation then the forthcoming results and conclusions could also be applied to the case of homogeneous cosmic birefringence.

\paragraph{Bias on parity violation--} Following the approach of Ref. \cite{keating_etal_2013}, the miscalibration angle can be fitted by minimizing the following $\chi^2$ (here generalized to the six angular power spectra):
\begin{eqnarray}
	\chi^2&=&\ds\sum_{A,A'}\sum_{b,b'}\left(C^{A~\mathrm{(th)}}_b-C^{A~\mathrm{(obs)}}_b\right)^\dag\times\left[\mathbf{\Sigma}^{-1}\right]^{A,A'}_{b,b'} \nonumber \\
	&\times&\left(C^{A'~\mathrm{(th)}}_{b'}-C^{A'~\mathrm{(obs)}}_{b'}\right),
\end{eqnarray}
with $C^{A~\mathrm{(th)}}_b$ the theoretically predicted angular power spectra, considered as functions of the set of parameters enlarged to $\theta_i=(r_{(+)},r_{(-)},\Delta\psi)$, and $C^{A~\mathrm{(obs)}}_b$ the reconstructed angular power spectra. The error bars on the reconstructed values of the parameters is the Fisher information matrix at the peak of the likelihood and is given by Eq. (\ref{eq:fishermat}), assuming that the estimated power spectra $C^{A~\mathrm{(obs)}}_b$ are unbiased.

The first effect of a miscalibration on the detection of parity violation would be to bias the measurement of $r_{(-)}$ if such rotation is not taken into account in the modelized $C_\ell$'s. Assuming that $C^{A~\mathrm{(th)}}_b$ is not rotated by the miscalibration angle though the observed spectra are, the recovered, and therefore {\it biased}, value of $r_{(-)}$, noted $r^{\mathrm{(bias)}}_{(-)}$, is obtained by minimizing the $\chi^2$ :
\begin{eqnarray}
	0&=&\ds\sum_{A,A'}\sum_{b,b'}\left(\frac{\partial C^{A~\mathrm{(th)}}_b}{\partial r_{(-)}}\right)^\dag\times\left[\mathbf{\Sigma}^{-1}\right]^{A,A'}_{b,b'} \\
	&\times&\left(C^{A'~\mathrm{(th)}}_{b'}-C^{A'~\mathrm{(obs)}}_{b'}\right). \nonumber
\end{eqnarray}
In the above, $C^{A~\mathrm{(obs)}}_{b}$ is fixed by the targeted values $\bar{\theta}_i=(\bar{r}_{(+)},\bar{r}_{(-)},\bar{\Delta}\psi)$ while $C^{A~\mathrm{(th)}}_b$ is a function of $r_{(\pm)}$ only, {\it i.e.} $C^{A~\mathrm{(th)}}_b=\tilde{C}^A_b$. The uncertainties on the reconstructed value of $r_{(-)}$ is derived from the Fisher matrix where $(\partial C^{A~\mathrm{(th)}}_b/\partial r_{(-)})=(\partial \tilde{C}^A_b/\partial r_{(-)})$.
\begin{figure}
\begin{center}
	\includegraphics[scale=0.5]{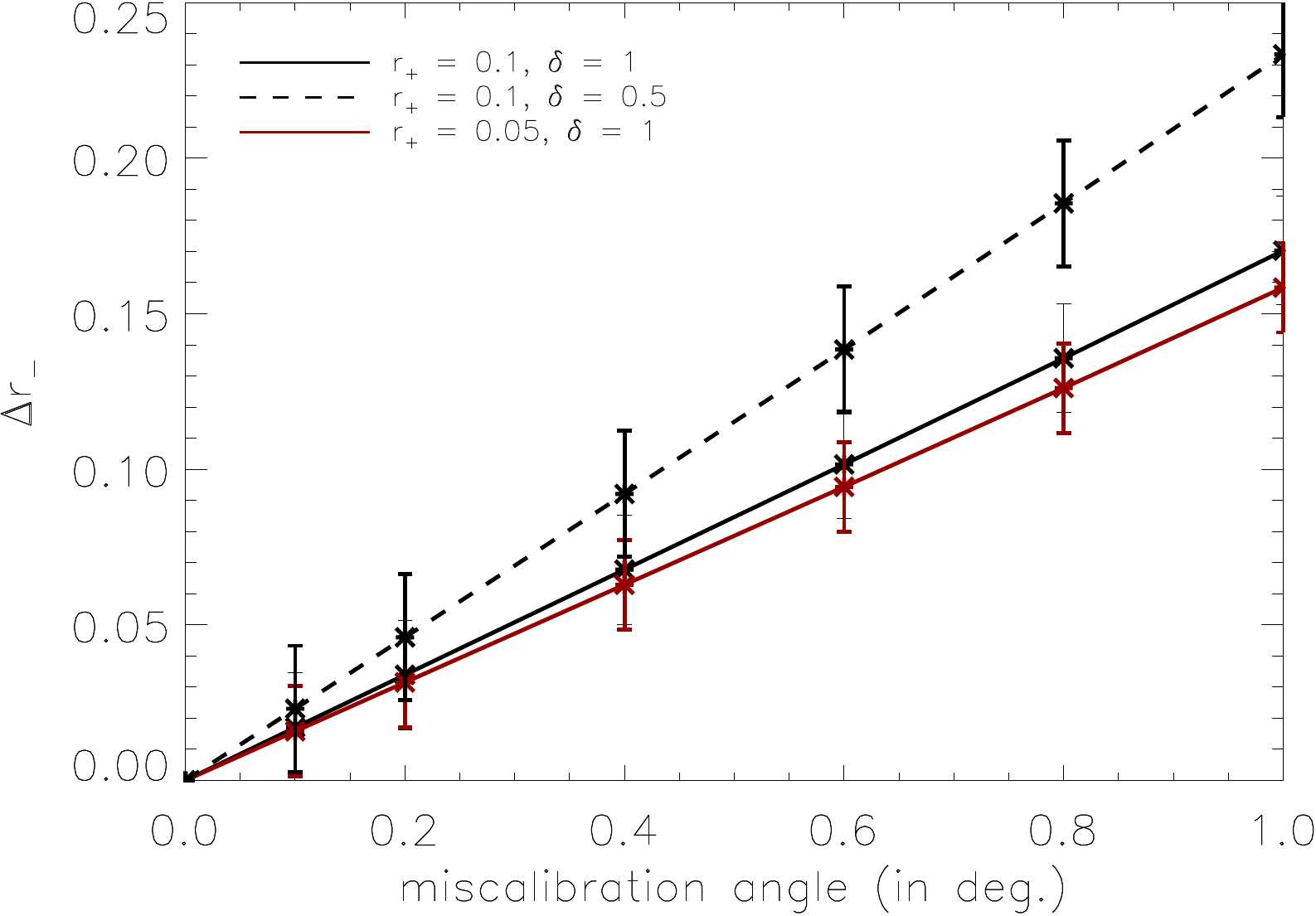}
	\caption{Bias on the reconstruction of $r_{(-)}$ if the miscalibration angle, $\Delta\psi$, is not taken into account in the parameters estimation. For $\Delta\psi\geq0.1$ degree, the bias is sufficiently high to lead to a false detection of parity violation.}
	\label{fig:biasrm}
\end{center}
\end{figure}

The biases, $\Delta r_{(-)}=(r^{\mathrm{(bias)}}_{(-)}-\bar{r}_{(-)})$, and their associated error bars are depicted in Fig. \ref{fig:biasrm} as a function of $\Delta\psi$ and for different input values of $r_{(-)}$. It shows that for $\Delta\psi \lesssim 0.1$ degree, the measured $r_{(-)}$ is compatible with the input value, $\bar{r}_{(-)}$, within the 1$\sigma$ uncertainty. For higher values of $\Delta\psi$, the bias would translate into a false detection of parity violation. We mention that most of the bias comes from the $EB$ spectrum since its intrinsic CMB contribution is rapidly dominated by spurious power spectra due to the miscalibration, {\it i.e.} $\Delta C^{EB}_\ell\geq \tilde{C}^{EB}_\ell$ for $\Delta\psi\geq0.1$ degree. The main contribution for small angles to the $EB$ power spectrum is $\tilde{C}^{EB~\mathrm{(rot)}}_\ell\simeq\tilde{C}^{EB}_\ell+2(\Delta\psi)\tilde{C}^{EE}_\ell$ with the first term being the intrinsic $EB$ correlations and the second term the spurious $EB$ correlation as induced by miscalibration of the orientation of the polarisers. This means that positive-valued $\Delta\psi$ leads to a positive bias while negative values of $\Delta\psi$ leads to a negative bias.

\paragraph{Statistical uncertainties with mode-counting--} The proper approach consists in fitting for the three parameters $\theta_i=(r_{(+)},r_{(-)},\Delta\psi)$ by minimizing the $\chi^2$. The final estimate will be unbiased and the uncertainties are given by the Fisher information matrix assuming that the theoretical power spectra are appropriately modeled, {\it i.e.} $C^{A~\mathrm{(th)}}_b=\tilde{C}^{A~\mathrm{(rot)}}_b$. As $r_{(-)}$ and $\Delta\psi$ could be degenerated, one should expect the marginalized error bars on $r_{(-)}$ to be enlarged. 

Using the mode-counting for estimating the covariance on the reconstructed angular power spectra, our numerical investigations show that the signal-to-noise ratio on $r_{(-)}$, for $r_{(\pm)}$ ranging from $0.004$ to $0.2$ and $\Delta\psi$ varying from 0.1 to 1 degree, are only degraded by a factor $\sim10^{-5}$ as compared to the case without such miscalibration angle, meaning that the {\it joined} reconstruction of $\Delta\psi$ with $r_{(-)}$ only marginally affects the detection of parity violation. This is in perfect agreement with the results obtained in {\it e.g.} Ref. \cite{gluscevic_etal_2010} where they consider both parity violation in the primordial universe and homogeneous cosmic birefringence. However, this conclusion is only valid assuming the mode-counting expressions for the covariance on the estimated angular power spectra. In the appendix \ref{appb}, we prove the following statement : assuming that i) the covariance matrix on the estimated angular power spectra, $\mathbf{\Sigma}$, is given by the mode-counting expression, ii) the entire set of correlations between the six estimated angular power spectra is taken into account, and, iii) the covariance matrix $\mathbf{\Sigma}$ is dominated by the sampling variance, then it is shown that the sub-block $(r_{(\pm)},r_{(\pm)})$ of the Fisher information matrix is equal to the Fisher matrix as derived {\it without} any miscalibration angle, and that the correlations between $r_{(\pm)}$ and $\Delta\psi$ {\it does not} depend on the values of $\Delta\psi$. Theoretically speaking, the hypothesis of the sampling variance dominated regime is not met at small angular scales. However, most of the constraints on $r_{(\pm)}$ come from the largest angular scales where noise variance is negligible. As a consequence, supposing that uncertainties are dominated by sampling variance is a valid assumption {\it in practice}. Moreover, our numerical results shows that the parameters $r_{(\pm)}$ are poorly degenerated with $\Delta\psi$ for the range of values of $\Delta\psi$ here-considered, {\it i.e.} 
$$
\frac{\mathbf{F}_{r_{(\pm)},\Delta\psi}}{\sqrt{\mathbf{F}_{r_{(\pm)},r_{(\pm)}}~\mathbf{F}_{\Delta\psi,\Delta\psi}}}\sim10^{-6}-10^{-5}. 
$$
This means that in practice, the (symmetric) Fisher information matrix is well approximated by
$$
\mathbf{F}\simeq\left(\begin{array}{ccc}
	F_{r_{(+)},r_{(+)}} & F_{r_{(+)},r_{(-)}} & \epsilon_{(+)}f_{(+)}(r_{(+)}) \\
	\cdot & F_{r_{(-)},r_{(-)}} & \epsilon_{(-)}f_{(-)}(r_{(-)}) \\
	\cdot & \cdot & F_{\Delta\psi,\Delta\psi}
\end{array}\right),
$$
with $F_{r_{(\pm)},r_{(\pm)}}$ derived {\it without} a miscalibration error, as in Sec. \ref{sssec:modecount}, and $\epsilon_{(\pm)}\sim10^{-5}$. As a consequence, the signal-to-noise ratio as derived in the previous section, Sec. \ref{sssec:modecount}, remains relevant even with a non-vanishing miscalibration angle.

We have further checked the above argument by two numerical experiments. On the one hand, we significantly increased the noise level which breaks the hypothesis of being sampling-variance dominated. On the other hand, we only take the diagonal of the covariance matrix $\mathbf{\Sigma}$, which break the hypothesis of using all the information through the correlations between different angular power spectra. In both cases, we do observe that the signal-to-noise ratio on {\it e.g.} $r_{(-)}$ indeed decreases for higher values of $\Delta\psi$. We have also checked numerically that $\mathbf{F}_{r_{(\pm)},\Delta\psi}$ does not depend on the value of $\Delta\psi$.

\paragraph{Statistical uncertainties with the pure pseudospectrum approach--} 
We have finally investigated the effect of a joint reconstruction of the miscalibration angle, $\Delta\psi$, and $r_{(-)}$ in the context of the pure pseudospectrum estimation of the $C_\ell$'s. We consider the two cases $r_{(+)}=r_{(-)}=0.1$ and 0.2 and subsequently derive the signal-to-noise ratio on $r_{(-)}$ marginalized over both $r_{(+)}$ {\it and} $\Delta\psi$. If $\Delta\psi$ is indeed equal to zero, the detection of $r_{(-)}$ is only marginally affected since the signal-to-ratio ratio is degraded by a relative factor of $10^{-4}$. However, this signal-to-noise ratio is reduced for higher values of $\Delta\psi$.  For $\Delta\psi=0.1$ degree, (S/N)$_{r_{(-)}}$ is reduced to 5 for $r_{(+)}=r_{(-)}=0.2$ (to be compared to 5.46 without miscalibration) and to 2.96 for $r_{(+)}=r_{(-)}=0.1$ (to be compared to 3.67). This corresponds to a decrease by a factor $\sim1.09$. For $\Delta\psi=1$ degree, the reduction factor is $\sim2.4$, obtaining (S/N)$_{r_{(-)}}$=2.23 for $r_{(+)}=r_{(-)}=0.2$ and (S/N)$_{r_{(-)}}$=1.58 for $r_{(+)}=r_{(-)}=0.1$. This shows that the non-degeneracy between $r_{(+)}$ and $\Delta\psi$ (as mentionned in Ref. \cite{gluscevic_etal_2010}) is only valid in the context of the mode-counting expression for the covariance. 

We believe the fundamental reason for such a result is that pseudospectrum based estimators does not allow for accessing to the whole set of correlations between the six estimated angular power spectra, due to the joint effect of mode-mixing and binning. Since keeping track of all the correlations was one of the mandatory assumptions used in App. \ref{appb}, this probably explains why in the more realistic case of pure pseudospectrum reconstruction of the $C_\ell$'s, non-vanishing $\Delta\psi$ then impacts on the significance of the estimation of $r_{(-)}$.

\paragraph{Inhomogeneous cosmic birefringence--} As previously underlined, the impact of miscalibrating the orientation of the polarized detectors is identical to the cosmological effect of {\it homogeneous} cosmic birefringence. However, such cosmic birefringerence can also have an {\it inhomogeneous} contribution. This could be the case if {\it e.g.} a scalar field coupled to the fermion current (therefore generating CPT violation) also exhibits some inhomogeneities in its energy density, as it should be since such a scalar field inevitably evolves in a {\it perturbed} FLRW space-time \cite{li_zhang_2008,zhao_li_2014}. In that case the rotation angle due to cosmic birefringence splits into an homogeneous part, $\alpha$, and an inhomogeneous part, $\delta\alpha(\vec{n})$, with $\delta\alpha\ll1$. This allows for performing a Taylor expansion to infer the impact of that inhomogeneous sector on the CMB angular power spectra. 

Considering this additional contribution and the presence of primary $EB$ and $TB$ contribution, the observed angular power for {\it e.g.} $TB$ correlations will receive new contributions proportional to $(\sin(2\alpha)C^{TE}_{\ell'}+\cos(2\alpha)C^{TB}_{\ell'})\times \left<\delta\alpha^2\right>$. Similar terms arise for the $EB$ spectrum where $(\sin(2\alpha)C^{TE}_{\ell'}+\cos(2\alpha)C^{TB}_{\ell'})$ is replaced by $(\sin(4\alpha)(\tilde{C}^{EE}_{\ell'}-\tilde{C}^{BB}_{\ell'})/2+\cos(4\alpha)\tilde{C}^{EB}_{\ell'})$. (We refer the interested reader to Refs. \cite{li_zhang_2008,zhao_li_2014} for a more detailed computation and we only focus here on orders of magnitude.) Those corrections are of second order, being proportional to $\left<\delta\alpha^2\right>$. Following Ref. \cite{li_zhang_2008}, the amplitude of $\left<\delta\alpha^2\right>$ has been estimated to be of the order of $\sim10^{-3}$. This means that the biases derived by assuming a purely homogeneous birefringence may changed by a factor $\sim10^{-3}$ which is well within the statistical uncertainties, making our previously derived results still relevant for inhomogeneous cosmic birefingence.

\subsection{Detection of $r_{(-)}$: small-scale experiment}
\label{sec:smares}
Our results on the signal-to-noise for $r_{(-)}$ in the case of a small-scale experiment are summarized in Tab. \ref{tab:snrrmsmall}, assuming the mode-counting for the derivation of the covariance matrix. Clearly in this case, the measurement of such a parameter is unfeasible as (S/N)$_{r_{(-)}}<1.5$, even in the most optimistic case of $r_{(+)}=r_{(-)}=0.2$. This is because most of the constrains on $r_{(-)}$ comes from the largest angular scale which are unachievable for an experiment covering 1\% of the celestial sphere. 

Using the covariance as obtained from a pure pseudospectrum estimation of the angular power spectra only degrades the signal-to-noise ratio. For example, in the case of $r_{(+)}=r_{(-)}=0.1$ ({\it i.e.} $\delta=1$), we obtain (S/N)$_{r_{(-)}}=0.2$, as compared to 0.64 by using the mode-counting approach. 

For a parity-invariant primordial universe, {\it i.e.} $r_{(-)}=0$, the marginalized uncertainty on $r_{(-)}$ for $r_{(+)}=0.05$ is $\sigma_{r_{(-)}}=0.36$. Therefore, the $\delta$ parameter is greater than unity which is theoretically irrelevant, meaning that no significant upper bound on the level of parity violation can be established using datas from ongoing or forthcoming small-scale experiments.

\begin{table}
\begin{center}
\begin{tabular}{cc||c|c|c|c|c|c}
	& $r_{(+)}$ & $0.2$ & $0.1$ & $0.07$ & $0.05$ & $ 0.03$ & $0.007$ \\
	$r_{(-)}$ & & & & & & & \\
	\hline\hline
	$0.2$    & & 1.22   &         &             &               &               &                \\
	\hline
	$0.1$    & & 0.43   & 0.64 &             &               &               &                \\
	\hline
	$0.07$  & &  0.29  & 0.4  & 0.487    &               &               &              \\
	\hline
	$0.05$  & &  0.2   &0.28 & 0.326    & 0.38     & 		    &  	     \\
	\hline
	$0.03$  & &  0.12  &0.16  & 0.188  & 0.216 & 0.27 &               \\
	\hline
	$0.007$ & & 0.03  &0.037& 0.043 & 0.049 & 0.06  &  0.1   \\
	\hline
\end{tabular}
	\caption{Signal-to-noise on $r_{(-)}$ for different values of $r_{(+)}$ in the case of small-scale (ballon-borne or ground-based) experiments, and using a mode-counting expression for the error bars on the angular power spectra reconstruction.} 
	\label{tab:snrrmsmall}
\end{center}
\end{table}

\section{Conclusion and discussion}
\label{sec:concl}

In this paper, we investigate the constraints that could be set on chiral gravity from the detection of the CMB $TB$ and $EB$ correlations, taking into account statistical uncertainties as incurred by pure pseudospectrum reconstruction of the CMB angular power spectra and considering the impact of miscalibrating the orientation of the polarized detectors. (We stress that all the constraints we have set are for positive valued $r_{(-)}$. They however equally apply to negative values of $r_{(-)}$ as in practice, the derived constraints are for $\left|r_{(-)}\right|$.)

We have shown that such a detection of parity violation leading to non zero $C^{TB(EB)}_\ell$ is beyond the scope of forthcoming small-scale measurements of CMB polarized anisotropies. Even in the most optimistic case of 100\% of parity violation and a tensor-to-scalar ratio of 0.2, and underestimating the uncertainties by using a mode-counting approach, the signal-to-noise ratio on the amplitude of parity asymmetric tensor mode is only of $\sim1.2$, and it rapidly diminishes to values smaller than unity for smaller values of the tensor-to-scalar ratio, $r_{(+)}$, or a smaller percentage of parity violation. This is because most of the constraints come from the largest angular scales which cannot be measured with enough significance by those experiments. Moreover, even in the case of vanishing $TB$ and $EB$ cross-correlations, the statistical uncertainties on their reconstruction via pure pseudospectrum estimators lead to an upper bound of the level of parity violation of more than 100\% at 95\% CL. Since this level is theoretically bounded from above at 100\%, this means that no significant constraint can be set on this type of parity violation using datas from ongoing or forthcoming small-scale experiments.

In the case of a potential satellite mission dedicated to primordial $B$-mode, we have shown that a detection with at least $2\sigma$ is possible for 100\% of parity violation and a tensor-to-scalar ratio of at least 0.05. A 1$\sigma$ detection is still achieved for 50\% of parity violation and a tensor-to-scalar ratio of at least 0.05 and a $2.5\sigma$ detection would be possible for $r_{(+)}=0.2$. We found that by a measurement of vanishing $TB$ and $EB$ angular power spectra using pure pseudospectrum estimators, the level of parity violation is bounded from above: $\left|\delta\right|\leq0.92$ at 95\% CL. We have also shown that for such an experimental configuration where sampling variance is dominating at the largest scales -- precisely those scales which allows for constraining parity violation --, the impact of self-calibrating the miscalibration angle could have a significant impact on the final estimation of the level of parity violation, the reported signal-to-noise ratio being degraded by a factor of $\sim1.09$ for a miscalibration angle of 0.1 degree and, more significantly, by a factor $\sim2.4$ for an angle of 1 degree. In this very last case, a $2\sigma$ detection of parity violation becomes possible only for $\delta=1$ and $r_{(+)}=0.2$. We stress that such an impact is revealed in the context of the pseudospectrum estimation of the angular power spectra. By making use of the na\"\i{v}e mode-counting expression for the covariance of the reconstructed $C_\ell$'s, it is formally shown that self-calibration of the orientation of the polarizers does not impact on the significance of the reconstruction of $r_{(-)}$ assuming that the covariance is dominated by sampling variance and that one have access to the entire set of cross-correlations between the six estimated angular power spectra. Nevertheless, this last assumption is broken by pseudospectrum estimators (because of mode-mixing and binning) leading to degeneracies between $r_{(-)}$ and $\Delta\psi$.

In the context of chiral gravity from the Ashtekar formulation of general relativity, the parameter $\delta$ amounting the level of parity breaking is related to the imaginary Barbero-Immirzi parameter via $\delta=2i\gamma/(1-\gamma^2)$ \cite{bethke_magueijo_2011a}. (We will restrict to the case of purely imaginary values of $\gamma$ though the formalism can be extended to the case of any arbitrary complex values of $\gamma$ \cite{bethke_magueijo_2011b}.) The (absolute) level of parity breaking is encoded in $\left|\delta\right|$ leading to $\left|\gamma\right|=\left(1\pm\sqrt{1-\delta^2}\right)/\left|\delta\right|$. Considering the statistical error bars from a pseudospectrum reconstruction of the $C_\ell$'s, detecting $\left|\gamma\right|=1$ is possible using datas from a satellite mission with a statistical significance ranging from $2.3\sigma$ to $5.4\sigma$ for a tensor-to-scalar ratio ranging from $0.05$ to $0.2$, respectively. Assuming a detection of $\left|\delta\right|=0.5$ translates into a detectable value of $\gamma=0.26$ or $\gamma=3.73$, meaning that such a form of chiral gravity is detectable with CMB polarized anisotropies if $0.26\leq\left|\gamma\right|\leq3.75$. The significance of that detection for a future satellite mission ranges from $1.1\sigma$ to $2.5\sigma$ for a tensor-to-scalar ratio of 0.05 and 0.2. Detecting $TB$ and $EB$ angular power spectra which are consistant with zero leads to an upper bound on the parity violation level $\left|\delta\right|\leq0.92$ at 95\% CL for $r_{(+)}=0.05$ and $\delta\leq0.39$ at 95\% CL for $r_{(+)}=0.2$. This would mean that $0.66\leq\left|\gamma\right|\leq1.5$ is excluded at 95\% CL for $r_{(+)}=0.05$ (the exclusion range at 68\% CL would be $0.24\leq\left|\gamma\right|\leq4.1$), and that $0.2\leq\left|\gamma\right|\leq4.9$ is excluded at 95\% CL for $r_{(+)}=0.2$ (the exclusion range at 68\% CL would be $0.098\leq\left|\gamma\right|\leq10.1$).

In the context of a pseudoscalar inflaton, the amount of parity violation is given by \cite{cook_sorbo}
\begin{equation}
	\left|\delta\right|=\frac{8.6\times10^{-7}\left(\frac{H^2}{2M^2_\mathrm{Pl}}\frac{e^{4\pi\xi}}{\xi^6}\right)}{1+8.6\times10^{-7}\left(\frac{H^2}{2M^2_\mathrm{Pl}}\frac{e^{4\pi\xi}}{\xi^6}\right)} \nonumber
\end{equation}
and
\begin{equation}
	r_{(+)}=8.1\times10^{7}\left(\frac{H^2}{M^2_\mathrm{Pl}}\right)\left[1+8.6\times10^{-7}\left(\frac{H^2}{2M^2_\mathrm{Pl}}\frac{e^{4\pi\xi}}{\xi^6}\right)\right]. \nonumber
\end{equation}
For a given value of $r_{(+)}$, one can express $\frac{H^2}{2M^2_\mathrm{Pl}}$ as a function of $\frac{e^{4\pi\xi}}{\xi^6}$ and plug it into $\left|\delta\right|$. Following Ref. \cite{cook_sorbo}, one introduces the parameter $\tilde{X}=e^{2\pi\xi}/\xi^3$ which is related to $r_{(+)}$ and $\delta$ via
\begin{equation}
	\tilde{X}=\left(\frac{1.37\times10^7}{\sqrt{r_{(+)}}}\right)\sqrt{\left(\frac{\left|\delta\right|}{1-\left|\delta\right|}\right)\left(1+\frac{\left|\delta\right|}{1-\left|\delta\right|}\right)}. \nonumber
\end{equation}
As compared to Ref. \cite{cook_sorbo}, $\tilde{X}$ is related to their $X$ parameter via $X=\epsilon\times\tilde{X}$ with $\epsilon$ the first slow-roll parameter. For $r_{(+)}=0.05$, one obtains the following upper bound on $\tilde{X}$: $\tilde{X}\leq73\times10^7$ at 95\% CL. For $r_{(+)}=0.2$, this upper bound is strengthened to $\tilde{X}\leq3\times10^{7}$ at 95\% CL. This has to be compared to the upper bound reported in Ref. \cite{cook_sorbo}: $\tilde{X}\leq6\times10^7$ at 95\% CL,  using the upper bound set by Planck on primordial non-gaussianities, $f_\mathrm{NL}<150$ \cite{planck_gauss}. This means that constraining such models with $TB$ and $EB$ is on par with the constraints that can be set with measurements of non-gaussianities assuming a rather high value of $r_{(+)}$.

\acknowledgments
The authors would like to warmly thank J. Lesgourgues for helpful discussions about the {\sc class} code as well as J. Peloton and R. Stompor for suggesting to look at the impact of a miscalibration angle. J.G. is grateful to T. Cailleteau, L. Smolin and F. Vidotto for enlightenning discussions on chiral gravity at the linear level. A.F. is also grateful to G. Hurier for his help on bias issues. This research used resources of the National Energy Research Scientific Computing Center, which is supported by the Office of Science of the U.S. Department of Energy under Contract No. DE-AC02-05CH11231. Some of the results in this paper have been derived using {\sc class}~\cite{class}, {\sc s$^2$hat} \cite{s2hat,pures2hat,hupca_etal_2010,szydlarski_etal_2011} and {\sc healpix}~\cite{gorski_etal_2005} software packages.


\begin{appendix}
\section{Lensed angular power spectra with primary $TB$ and $EB$ correlations}
\label{app:lensing}
Weak lensing of the primary anisotropies remaps the primary anisotropies by a displacement fields $\delta \vec{n}=\vec{\nabla}\phi$ with $\nabla$ the covariant derivative on the sphere and $\phi$ the projected potential of the large scale structures.  By denoting $\tilde{X}$ the {\it lensed} CMB anisotropies (we keep untilted notations for the primary CMB anisotropies), this translates into $T(\vec{n})\to\tilde{T}(\vec{n})=T(\vec{n}+\delta\vec{n})$ and $P_{\pm}(\vec{n})\to\tilde{P}_{\pm}(\vec{n})=P_{\pm}(\vec{n}+\delta\vec{n})$. This displacement field is supposed to be small in amplitude and one can therefore perform a Taylor expansion up to second order of the lensed CMB anisotropies around their unlensed value. With such a Taylor expansion, it is shown that the remapping of CMB primary fluctuations appears as a reshuffling of the multipoles $X_{\ell m}$ according to some convolution kernel involving the harmonic decomposition of the projected potential, $\phi_{\ell m}$ \cite{hu_2000}. This reshuffling acts between different $\ell$-multipoles but also between different polarization modes as lensed $B$-modes receive contribution from primary $E$-modes and vice versa. For temperature, it reads:
\begin{widetext}
\begin{eqnarray}
	\tilde{T}_{\ell m}&=&T_{\ell m}+\underbrace{\displaystyle\sum_{\ell_1,m_1}\sum_{\ell_2,m_2}\phi_{\ell_1m_1}T_{\ell_2m_2}\times I^{\ell\ell_1\ell_2}_{mm_1 m_2}}_{T^{(1)}_{\ell m}}  +\underbrace{\frac{1}{2}\displaystyle\sum_{\ell_1,m_1}\sum_{\ell_2,m_2}\sum_{\ell_3,m_3}\phi_{\ell_1m_1}T_{\ell_2m_2}\phi^\star_{\ell_3m_3}\times J^{\ell\ell_1\ell_2\ell_3}_{mm_1m_2m_3}}_{T^{(2)}_{\ell m}}, \label{eq:Tshuff}
\end{eqnarray}
with
\begin{eqnarray}
	I^{\ell\ell_1\ell_2}_{mm_1 m_2}&=&\displaystyle\int_{4\pi}Y^\star_{\ell m}\left(\nabla_aY_{\ell_1m_1}\right)\left(\nabla^aY_{\ell_2m_2}\right)~d\vec{n}, \\
	J^{\ell\ell_1\ell_2\ell_3}_{mm_1m_2m_3}&=&\displaystyle\int_{4\pi}Y^\star_{\ell m}\left(\nabla_aY_{\ell_1m_1}\right)\left(\nabla_bY^\star_{\ell_3m_3}\right)\left(\nabla^a\nabla^bY_{\ell_2m_2}\right)~d\vec{n}.
\end{eqnarray}
Similarly for polarization, one obtains:
\begin{eqnarray}
\tilde{E}_{\ell m}&=&E_{\ell m}+\underbrace{\frac{1}{2}\displaystyle\sum_{\ell_1,m_1}\sum_{\ell_2,m_2}\phi_{\ell_1m_1}\left[E_{\ell_2m_2}\times {}_{(+)}I^{\ell\ell_1\ell_2}_{mm_1 m_2}+iB_{\ell_2m_2}\times {}_{(-)}I^{\ell\ell_1\ell_2}_{mm_1 m_2}\right]}_{E^{(1)}_{\ell m}}  \label{eq:Eshuff} \\
	&+&\underbrace{\frac{1}{4}\displaystyle\sum_{\ell_1,m_1}\sum_{\ell_2,m_2}\sum_{\ell_3,m_3}\phi_{\ell_1m_1}\phi^\star_{\ell_3m_3}\left[E_{\ell_2m_2}\times {}_{(+)}J^{\ell\ell_1\ell_2\ell_3}_{mm_1m_2m_3}+iB_{\ell_2m_2}\times {}_{(-)}J^{\ell\ell_1\ell_2\ell_3}_{mm_1m_2m_3}\right]}_{E^{(2)}_{\ell m}}, \nonumber 
	\\
\tilde{B}_{\ell m}&=&B_{\ell m}+\underbrace{\frac{1}{2}\displaystyle\sum_{\ell_1,m_1}\sum_{\ell_2,m_2}\phi_{\ell_1m_1}\left[B_{\ell_2m_2}\times {}_{(+)}I^{\ell\ell_1\ell_2}_{mm_1 m_2}-iE_{\ell_2m_2}\times {}_{(-)}I^{\ell\ell_1\ell_2}_{mm_1 m_2}\right]}_{B^{(1)}_{\ell m}}  \label{eq:Bshuff} \\
	&+&\underbrace{\frac{1}{4}\displaystyle\sum_{\ell_1,m_1}\sum_{\ell_2,m_2}\sum_{\ell_3,m_3}\phi_{\ell_1m_1}\phi^\star_{\ell_3m_3}\left[B_{\ell_2m_2}\times {}_{(+)}J^{\ell\ell_1\ell_2\ell_3}_{mm_1m_2m_3}-iE_{\ell_2m_2}\times {}_{(+)}J^{\ell\ell_1\ell_2\ell_3}_{mm_1m_2m_3}\right]}_{B^{(2)}_{\ell m}}, \nonumber
\end{eqnarray}
with
\begin{eqnarray}
	{}_{(\pm)}I^{\ell\ell_1\ell_2}_{mm_1 m_2}&=&\displaystyle\int_{4\pi}\left[{}_{2}Y^\star_{\ell m}\left(\nabla_aY_{\ell_1m_1}\right)\left(\nabla^a{}_{2}Y_{\ell_2m_2}\right)\pm{}_{-2}Y^\star_{\ell m}\left(\nabla_aY_{\ell_1m_1}\right)\left(\nabla^a{}_{-2}Y_{\ell_2m_2}\right)\right]d\vec{n}, \\
	{}_{(\pm)}J^{\ell\ell_1\ell_2\ell_3}_{mm_1m_2m_3}&=&\displaystyle\int_{4\pi}\left[{}_{2}Y^\star_{\ell m}\left(\nabla_aY_{\ell_1m_1}\right)\left(\nabla_bY^\star_{\ell_3m_3}\right)\left(\nabla^a\nabla^b{}_{2}Y_{\ell_2m_2}\right)\pm{}_{-2}Y^\star_{\ell m}\left(\nabla_aY_{\ell_1m_1}\right)\left(\nabla_bY^\star_{\ell_3m_3}\right)\left(\nabla^a\nabla^b{}_{-2}Y_{\ell_2m_2}\right)\right]d\vec{n}. \nonumber
\end{eqnarray}
\end{widetext}
The kernels $I^{\ell\ell_1\ell_2}_{mm_1m_2}$ and $J^{\ell\ell_1\ell_2\ell_3}_{mm_1m_2m_3}$ have properties which greatly simplifies the forthcoming computations.

The final derivation of the lensed angular power spectrum is obtained by considering the correlators $\tilde{C}^{XZ}_\ell=\left[\left<\tilde{X}_{\ell m}\tilde{Z}^\star_{\ell m}\right>+\left<\tilde{X}^\star_{\ell m}\tilde{Z}_{\ell m}\right>\right]/2$, and making use of the statistical isotropy of the primary fluctuations, {\it i.e.} $\left<{X}_{\ell m}{Z}^\star_{\ell' m'}\right>=C^{XZ}_\ell~\delta_{\ell\ell'}\delta_{mm'}$ and $\left<{\phi}_{\ell m}{\phi}^\star_{\ell' m'}\right>=C^{\phi\phi}_\ell~\delta_{\ell\ell'}\delta_{mm'}$ ; following Ref. \cite{hu_2000}, we also consider that the projected potential causing the deflection of CMB photons is not correlated to the primary anisotropies of CMB and we neglect any curl-like contribution to the deflection field. 

\subsection{Properties of $I^{\ell\ell_1\ell_2}_{mm_1m_2}$ and $J^{\ell\ell_1\ell_2\ell_3}_{mm_1m_2m_3}$}
\subsubsection{First property} 
The first important property is that $I^{\ell\ell_1\ell_2}_{mm_1 m_2}$ is an 'even' quantity:
\begin{equation}
	\left\{\begin{array}{l}
		I^{\ell\ell_1\ell_2}_{mm_1 m_2}=0~~~\mathrm{for}~(\ell+\ell_1+\ell_2)=2n+1, \\
		\\
		I^{\ell\ell_1\ell_2}_{mm_1 m_2}\neq0~~~\mathrm{for}~(\ell+\ell_1+\ell_2)=2n. \\
	\end{array}\right.
\end{equation}

As shown in Refs. \cite{hu_2000,goldberg_1999}, $I^{\ell\ell_1\ell_2}_{mm_1 m_2}$ is rewritten as a function of the Gaunt integral
\begin{eqnarray}
	I^{\ell\ell_1\ell_2}_{mm_1 m_2}&=&\frac{1}{2}\left[\ell_1(\ell_1+1)+\ell_2(\ell_2+1)-\ell(\ell+1)\right] \nonumber \\
	&\times&\displaystyle\int_{4\pi}Y^\star_{\ell m}(\vec{n})Y_{\ell_1m_1}(\vec{n})Y_{\ell_2m_2}(\vec{n})~d\vec{n}.
\end{eqnarray}
The second line of the above is precisely the Gaunt integral, $\mathcal{G}$, which is re-expressed as a function of the Wigner-$3j$'s, {\it i.e.}
\begin{eqnarray}
	\mathcal{G}&=&(-1)^m\sqrt{\frac{(2\ell+1)(2\ell_1+1)(2\ell_2+1)}{4\pi}} \\
	&\times&\left(\begin{array}{ccc}
			\ell & \ell_1 & \ell_2 \\
			-m & m_1 &m_2
		\end{array}\right)\left(\begin{array}{ccc}
			\ell & \ell_1 & \ell_2 \\
			0 & 0 &0
		\end{array}\right). \nonumber
\end{eqnarray}
Since
\begin{equation}
	\left(\begin{array}{ccc}
			\ell & \ell_1 & \ell_2 \\
			m & m_1 &m_2
		\end{array}\right)=(-1)^{\ell+\ell_1+\ell_2}	\left(\begin{array}{ccc}
			\ell & \ell_1 & \ell_2 \\
			-m & -m_1 &-m_2
		\end{array}\right) 
\end{equation}
it is obvious that the Wigner-$3j$ is vanishing for $m=m_1=m_2=0$ for $\ell+\ell_1+\ell_2=2n+1$. As a consequence, $I^{\ell\ell_1\ell_2}_{mm_1 m_2}$ is also equal to zero for odd values of $\ell+\ell_1+\ell_2$.

\subsubsection{Second and third properties} 
The second and third important properties are that ${}_{(+)}I^{\ell\ell_1\ell_2}_{mm_1 m_2}$ is an 'even' quantity, and, ${}_{(-)}I^{\ell\ell_1\ell_2}_{mm_1 m_2}$ is an 'odd' quantity, {\it i.e.}:
\begin{equation}
	\left\{\begin{array}{l}
		{}_{(+)}I^{\ell\ell_1\ell_2}_{mm_1 m_2}=0~~~\mathrm{for}~(\ell+\ell_1+\ell_2)=2n+1, \\
		\\
		{}_{(+)}I^{\ell\ell_1\ell_2}_{mm_1 m_2}\neq0~~~\mathrm{for}~(\ell+\ell_1+\ell_2)=2n. \\
	\end{array}\right.
\end{equation}
and
\begin{equation}
	\left\{\begin{array}{l}
		{}_{(-)}I^{\ell\ell_1\ell_2}_{mm_1 m_2}\neq0~~~\mathrm{for}~(\ell+\ell_1+\ell_2)=2n+1, \\
		\\
		{}_{(-)}I^{\ell\ell_1\ell_2}_{mm_1 m_2}=0~~~\mathrm{for}~(\ell+\ell_1+\ell_2)=2n. \\
	\end{array}\right.
\end{equation}

Those two properties are simply proved by noticing that \cite{hu_2000}
\begin{eqnarray}
	&&\displaystyle\int_{4\pi}{}_{2}Y^\star_{\ell m}\left(\nabla_aY_{\ell_1m_1}\right)\left(\nabla^a{}_{2}Y_{\ell_2m_2}\right)d\vec{n}= \\
	&&(-1)^{\ell+\ell_1+\ell_2}\displaystyle\int_{4\pi}{}_{-2}Y^\star_{\ell m}\left(\nabla_aY_{\ell_1m_1}\right)\left(\nabla^a{}_{-2}Y_{\ell_2m_2}\right)d\vec{n}. \nonumber
\end{eqnarray}
From that, it is obvious that ${}_{(+)}I^{\ell\ell_1\ell_2}_{mm_1 m_2}=0$ for odd values of $\ell+\ell_1+\ell_2$ while ${}_{(-)}I^{\ell\ell_1\ell_2}_{mm_1 m_2}=0$ for even values of $\ell+\ell_1+\ell_2$.

This can also be seen from the explicit expressions of ${}_{(\pm)}I^{\ell\ell_1\ell_2}_{mm_1 m_2}$ as functions of the Wigner-$3j$ since
\begin{eqnarray}
	{}_{(\pm)}I^{\ell\ell_1\ell_2}_{mm_1 m_2}&=&\frac{1}{2}\left[\ell_1(\ell_1+1)+\ell_2(\ell_2+1)-\ell(\ell+1)\right] \nonumber \\
	&\times&(-1)^{m+2}\sqrt{\frac{(2\ell+1)(2\ell_1+1)(2\ell_2+1)}{4\pi}} \nonumber \\
	&\times&\left(\begin{array}{ccc}
			\ell & \ell_1 & \ell_2 \\
			-m & m_1 &m_2
		\end{array}\right) \\ 
	&\times&\left[\left(\begin{array}{ccc}
			\ell & \ell_1 & \ell_2 \\
			-2 & 0 &2
		\end{array}\right)\pm\left(\begin{array}{ccc}
			\ell & \ell_1 & \ell_2 \\
			2 & 0 &-2
		\end{array}\right)\right]. \nonumber
\end{eqnarray}
From the symmetry of the Wigner-$3j$'s involved in the last line, it is clear that ${}_{(+/-)}I^{\ell\ell_1\ell_2}_{mm_1 m_2}=0$ for odd(even) values of $\ell+\ell_1+\ell_2$.

\subsubsection{Fourth property} 
The last useful property is
\begin{equation}
	\displaystyle\sum_{m_1}{}_{(-)}J^{\ell\ell_1\ell\ell_1}_{mm_1mm_1}=0.
\end{equation}
To prove it, one shoud first notice that (see Eqs. (59) and (60) of \cite{hu_2000})
\begin{eqnarray}
	\displaystyle\sum_{m_1}\nabla_aY_{\ell_1m_1}\nabla_bY^\star_{\ell_1m_1}&=&\frac{\ell_1(\ell_1+1)}{2}\frac{2\ell_1+1}{4\pi}\left[(\vec{m}_+)_a(\vec{m}_-)_b\right. \nonumber \\
	&+&\left.(\vec{m}_-)_a(\vec{m}_+)_b\right], \label{eq:propa}
\end{eqnarray}
and
\begin{eqnarray}
	&&\left[(\vec{m}_+)_a(\vec{m}_-)_b+(\vec{m}_-)_a(\vec{m}_+)_b\right]\nabla^a\nabla^b{}_{\pm s}Y_{\ell m}= \nonumber \\
	&&-\left[\ell(\ell+1)-s^2\right]{}_{\pm2}Y_{\ell m}. \label{eq:propb}
\end{eqnarray}
with $\vec{m}_{\pm}=(\vec{e}_\theta\mp i\vec{e}_\varphi)/\sqrt{2}$. From those two expressions, one easily derive that
\begin{eqnarray}
	\displaystyle\sum_{m_1}{}_{(-)}J^{\ell\ell_1\ell\ell_1}_{mm_1mm_1}&=&F(\ell,\ell_1)\displaystyle\int_{4\pi}\left[{}_{2}Y^\star_{\ell m}{}_{2}Y_{\ell m}\right. \\
	&-&\left.{}_{-2}Y^\star_{\ell m}{}_{-2}Y_{\ell m}\right]d\vec{n}, \nonumber
\end{eqnarray}
with $F(\ell,\ell_1)$ a numerical factor depending on $\ell$ and $\ell_1$. The integral in the right-hand side of the above expression is vanishing since the ${}_{\pm2}Y_{\ell m}$'s forms an orthonormal basis on the celestial sphere.

\subsection{Temperature power spectrum}
The case of $TT$ correlation is rather straighforward: it only consists of reshuffling $\ell$-multipoles without any contribution from primary power spectra but $C^{TT}_\ell$. The lensed power spectrum is therefore not affected by non-vanishing $TB$ and $EB$ and reads:
\begin{eqnarray}
	\tilde{C}^{TT}_\ell&=&\left[1+\displaystyle\sum_{\ell_3}C^{\phi\phi}_{\ell_3}\sum_{m_3}\mathrm{Re}\left[J^{\ell\ell_3\ell\ell_3}_{mm_3mm_3}\right]\right]C^{TT}_\ell \nonumber \\
	&+&\displaystyle\sum_{\ell_1,\ell_2}C^{TT}_{\ell_2}C^{\phi\phi}_{\ell_1}\sum_{m_1,m_2}\left|I^{\ell\ell_1\ell_2}_{mm_1m_2}\right|^2.
\end{eqnarray}
From the expression of $I^{\ell\ell_1\ell_2}_{mmm_1m_2}$ as a function of the Wigner-$3j$'s and using that
\begin{equation}
	\displaystyle\sum_{m_1,m_2}\left(\begin{array}{ccc}
			\ell & \ell_1 & \ell_2 \\
			m & m_1 &m_2
		\end{array}\right)\left(\begin{array}{ccc}
			\ell' & \ell_1 & \ell_2 \\
			m' & m_1 &m_2
		\end{array}\right)=\frac{\delta_{\ell\ell'}\delta_{mm'}}{2\ell+1}, \label{eq:orthowig}
\end{equation}
it is easily shown that
\begin{eqnarray}
		F^{T}_{\ell\ell_1\ell_2}&=&\displaystyle\sum_{m_1,m_2}\left|I^{\ell\ell_1\ell_2}_{mmm_1m_2}\right|^2 \\
		&=&\frac{1}{4}\left[\ell_1(\ell_1+1)+\ell_2(\ell_2+1)-\ell(\ell+1)\right]^2  \nonumber \\
	&\times&{\frac{(2\ell_1+1)(2\ell_2+1)}{4\pi}}\left(\begin{array}{ccc}
			\ell & \ell_1 & \ell_2 \\
			0 & 0 &0
		\end{array}\right)^2. \nonumber
\end{eqnarray}
For the second term, one makes use of Eqs. (\ref{eq:propa}) and (\ref{eq:propb}) to show that
\begin{eqnarray}
	\displaystyle\sum_{m_3}J^{\ell\ell_3\ell\ell_3}_{mm_3mm_3}&=&-\frac{1}{2}\ell_3(\ell_3+1)\ell(\ell+1)\frac{2\ell_3+1}{4\pi} \nonumber \\
	&\times&\displaystyle\int_{4\pi}Y^\star_{\ell m}Y_{\ell m}~d\vec{n} \nonumber \\
	&=&-\frac{1}{2}\ell_3(\ell_3+1)\ell(\ell+1)\frac{2\ell_3+1}{4\pi},
\end{eqnarray}
where the orthonormality of spherical harmonics has been used to derive the second line of the above equation. This leads to
\begin{eqnarray}
	R^{T}&=&\displaystyle\sum_{\ell_3}C^{\phi\phi}_{\ell_3}\sum_{m_3}\mathrm{Re}\left[J^{\ell\ell_3\ell\ell_3}_{mm_3mm_3}\right] \\
	&=&-\frac{1}{2}\ell(\ell+1)\displaystyle\sum_{\ell_3}\ell_3(\ell_3+1)\frac{2\ell_3+1}{4\pi}C^{\phi\phi}_{\ell_3}. \nonumber
\end{eqnarray}

\subsection{Temperature-polarization power spectrum}
We focus on the case of the $TB$ cross-correlation. This calculation is then easily adapted to the case of $TE$ by first, replacing $(C^{TB}_\ell)$ by $(C^{TE}_\ell)$, and, second, replacing $(C^{TE}_\ell)$ by $(-C^{TB}_\ell)$,. Let us first notice that 
\begin{eqnarray}
	\tilde{C}^{TB}_\ell&=&\left<T_{\ell m}B^\star_{\ell m}\right>+\left<T^{(1)}_{\ell m}B^{(1)\star}_{\ell m}\right> \\
	&+&\left<T_{\ell m}B^{(2)\star}_{\ell m}\right>+\left<T^{(2)}_{\ell m}B^\star_{\ell m}\right>. \nonumber
\end{eqnarray}
Each of this term are given by
\begin{widetext}
\begin{eqnarray}
	\left<T_{\ell m}B^\star_{\ell m}\right>&=&C^{TB}_\ell, \\
	\left<T^{(1)}_{\ell m}B^{(1)\star}_{\ell m}\right> &=&\frac{1}{2}\displaystyle\sum_{\ell_1,\ell_2}C^{\phi\phi}_{\ell_1}\left\{C^{TB}_{\ell_2}\sum_{m_1m_2}\mathrm{Re}\left[I^{\ell\ell_1\ell_2}_{mm_1m_2}{}_{(+)}I^{\ell\ell_1\ell_2~\star}_{mm_1m_2}\right]-C^{TE}_{\ell_2}\sum_{m_1m_2}\mathrm{Im}\left[I^{\ell\ell_1\ell_2}_{mm_1m_2}{}_{(-)}I^{\ell\ell_1\ell_2~\star}_{mm_1m_2}\right]\right\}  \label{eq:t1b1} \\
	\left<T_{\ell m}B^{(2)\star}_{\ell m}\right>&=&\frac{1}{4}C^{TB}_{\ell}\displaystyle\sum_{\ell_3}C^{\phi\phi}_{\ell_3}\sum_{m_3}\mathrm{Re}\left[{}_{(+)}J^{\ell\ell_3\ell\ell_3}_{mm_3mm_3}\right] +\frac{1}{4}C^{TE}_{\ell}\displaystyle\sum_{\ell_3}C^{\phi\phi}_{\ell_3}\sum_{m_3}\mathrm{Im}\left[{}_{(-)}J^{\ell\ell_3\ell\ell_3}_{mm_3mm_3}\right] \label{eq:t0b2} \\
	\left<T^{(2)}_{\ell m}B^\star_{\ell m}\right>&=&\frac{1}{2}C^{TB}_{\ell}\displaystyle\sum_{\ell_3}C^{\phi\phi}_{\ell_3}\sum_{m_3}\mathrm{Re}\left[J^{\ell\ell_3\ell\ell_3}_{mm_3mm_3}\right]
\end{eqnarray}
\end{widetext}
The above expressions can be simplified using the properties of the $I,~{}_{(\pm)}I$ and $J,~{}_{(\pm)}J$ kernels. First, the second line of Eq. (\ref{eq:t1b1}) proportional to $C^{TE}_\ell$ is zero. This is a consequence of the first and the third properties: for any triplet $(\ell,\ell_1,\ell_2)$, either $I^{\ell\ell_1\ell_2}_{mm_1m_2}$ or ${}_{(-)}I^{\ell\ell_1\ell_2}_{mm_1m_2}$ is zero and the product of the two is always vanishing. In other word, primary $TE$ does not contribute to $\left<T^{(1)}_{\ell m}B^{(1)\star}_{\ell m}\right>$, in perfect agreement with the analogous case of pseudospectrum estimators. (This is for that very same reason that $TE$ is not leaking into $TB$ in pseudospectrum estimator.) Second, the second line of Eq. (\ref{eq:t0b2}) proportionnal to $C^{TE}_\ell$ is also vanishing as a result of the fourth property. This means that even at second order, the {\it primary} $TE$ correlations do not contribute to the {\it lensed} $TB$ correlations.

The same strategy is adopted for the $TE$ angular power spectrum, and, similarly to $TB$, it appears that the primary $C^{TB}_\ell$ does not contribute to the lensed $\tilde{C}^{TE}_\ell$. Gathering all the terms, one finally obtains
\begin{eqnarray}
	\tilde{C}^{TE}_\ell=\left[1+R^{X}\right]C^{TE}_\ell+\displaystyle\sum_{\ell_1,\ell_2}F^{X}_{\ell\ell_1\ell_2}C^{\phi\phi}_{\ell_1}C^{TE}_{\ell_2}, \\
	\tilde{C}^{TB}_\ell=\left[1+R^{X}\right]C^{TB}_\ell+\displaystyle\sum_{\ell_1,\ell_2}F^{X}_{\ell\ell_1\ell_2}C^{\phi\phi}_{\ell_1}C^{TB}_{\ell_2},
\end{eqnarray}
with
\begin{eqnarray}
	F^{X}_{\ell\ell_1\ell_2}&=&\frac{1}{2}\displaystyle\sum_{m_1,m_2}\mathrm{Re}\left[I^{\ell\ell_1\ell_2}_{mmm_1m_2}{}_{(+)}I^{\ell\ell_1\ell_2\star}_{mmm_1m_2}\right], \\
	R^{X}&=&\frac{1}{4}\displaystyle\sum_{\ell_3}C^{\phi\phi}_{\ell_3}\sum_{m_3}\left\{2\mathrm{Re}\left[J^{\ell\ell_3\ell\ell_3}_{mm_3mm_3}\right]\right. \\
	&+&\left.\mathrm{Re}\left[{}_{(+)}J^{\ell\ell_3\ell\ell_3}_{mm_3mm_3}\right]\right\}. \nonumber
\end{eqnarray}
The first quantity is easily computed starting from the expressions of $I^{\ell\ell_1\ell_2}_{mmm_1m_2}$ and ${}_{(+)}I^{\ell\ell_1\ell_2}_{mmm_1m_2}$ as functions of the Wigner-$3j$ and using Eq. (\ref{eq:orthowig}) to get
\begin{eqnarray}
		F^{X}_{\ell\ell_1\ell_2}&=&\frac{1}{8}\left[\ell_1(\ell_1+1)+\ell_2(\ell_2+1)-\ell(\ell+1)\right]^2  \nonumber \\
	&\times&{\frac{(2\ell_1+1)(2\ell_2+1)}{4\pi}}\left(\begin{array}{ccc}
			\ell & \ell_1 & \ell_2 \\
			0 & 0 &0
		\end{array}\right) \\
	&\times&\left[\left(\begin{array}{ccc}
			\ell & \ell_1 & \ell_2 \\
			2 & 0 &-2
		\end{array}\right)\pm\left(\begin{array}{ccc}
			\ell & \ell_1 & \ell_2 \\
			-2 & 0 &2
		\end{array}\right)\right]. \nonumber
\end{eqnarray}
Using Eqs. (\ref{eq:propa}) and (\ref{eq:propb}) and the orthonormality of spin-$(\pm2)$ spherical harmonics, it is proved that
\begin{equation}
	\displaystyle\sum_{m_3}{}_{(+)}J^{\ell\ell_3\ell\ell_3}_{mm_3mm_3}=-\ell_3(\ell_3+1)\left[\ell(\ell+1)-4\right]\frac{2\ell_3+1}{4\pi}.
\end{equation}
This finally leads to
\begin{equation}
	R^{X}=-\frac{1}{2}\left[\ell(\ell+1)-2\right]\displaystyle\sum_{\ell_3}\ell_3(\ell_3+1)\frac{2\ell_3+1}{4\pi}C^{\phi\phi}_{\ell_3}.
\end{equation}

\subsection{Polarization power spectra}
For the case of polarized angular power spectra, we detailed our derivation of the lensed $EB$ cross-correlation which can then easily be adapted to the case of $EE$ and $BB$ angular power spectra. Following the same steps as the ones used for $\tilde{C}^{TB}_\ell$, we first note that
\begin{eqnarray}
	\tilde{C}^{EB}_\ell&=&\mathrm{Re}\left[\left<E_{\ell m}B^\star_{\ell m}\right>\right]+\mathrm{Re}\left[\left<E^{(1)}_{\ell m}B^{(1)\star}_{\ell m}\right>\right] \\
	&+&\mathrm{Re}\left[\left<E_{\ell m}B^{(2)\star}_{\ell m}\right>\right]+\mathrm{Re}\left[\left<E^{(2)}_{\ell m}B^\star_{\ell m}\right>\right]. \nonumber
\end{eqnarray}
Each of this term are given by
\begin{widetext}
\begin{eqnarray}
	\mathrm{Re}\left[\left<E_{\ell m}B^\star_{\ell m}\right>\right]&=&C^{EB}_\ell, \\
	\mathrm{Re}\left[\left<E^{(1)}_{\ell m}B^{(1)\star}_{\ell m}\right>\right] &=&\frac{1}{4}\displaystyle\sum_{\ell_1,\ell_2}C^{EB}_{\ell_2}C^{\phi\phi}_{\ell_1}\sum_{m_1m_2}\left(\left|{}_{(+)}I^{\ell\ell_1\ell_2~\star}_{mm_1m_2}\right|^2-\left|{}_{(-)}I^{\ell\ell_1\ell_2~\star}_{mm_1m_2}\right|^2\right) \label{eq:e1b1} \\
	&-&\frac{1}{4}\displaystyle\sum_{\ell_1,\ell_2}\left(C^{EE}_{\ell_2}-C^{BB}_\ell\right)C^{\phi\phi}_{\ell_1}\sum_{m_1m_2}\mathrm{Im}\left[{}_{(+)}I^{\ell\ell_1\ell_2}_{mm_1m_2}{}_{(-)}I^{\ell\ell_1\ell_2~\star}_{mm_1m_2}\right] \nonumber \\
	\mathrm{Re}\left[\left<E_{\ell m}B^{(2)\star}_{\ell m}\right>\right]&=&\frac{1}{4}C^{EB}_{\ell}\displaystyle\sum_{\ell_3}C^{\phi\phi}_{\ell_3}\sum_{m_3}\mathrm{Re}\left[{}_{(+)}J^{\ell\ell_3\ell\ell_3}_{mm_3mm_3}\right]+\frac{1}{4}C^{EE}_{\ell}\displaystyle\sum_{\ell_3}C^{\phi\phi}_{\ell_3}\sum_{m_3}\mathrm{Im}\left[{}_{(-)}J^{\ell\ell_3\ell\ell_3}_{mm_3mm_3}\right] \label{eq:e0b2} \\
	\mathrm{Re}\left[\left<E^{(2)}_{\ell m}B^\star_{\ell m}\right>\right]&=&\frac{1}{4}C^{EB}_{\ell}\displaystyle\sum_{\ell_3}C^{\phi\phi}_{\ell_3}\sum_{m_3}\mathrm{Re}\left[{}_{(+)}J^{\ell\ell_3\ell\ell_3}_{mm_3mm_3}\right]-\frac{1}{4}C^{BB}_{\ell}\displaystyle\sum_{\ell_3}C^{\phi\phi}_{\ell_3}\sum_{m_3}\mathrm{Im}\left[{}_{(-)}J^{\ell\ell_3\ell\ell_3}_{mm_3mm_3}\right]. \label{eq:e2b0}
\end{eqnarray}
\end{widetext}
From the fourth property, the second term in the right-hand side of the equations (\ref{eq:e0b2}) and (\ref{eq:e2b0}) are vanishing. Similarly, the second term in the right-hand side of Eq. (\ref{eq:e1b1}) is equl to zero as a result of the second and third properties. Therefore, both the primary $EE$ spectrum and the primary $BB$ spectrum does {\it not} contirbute to the lensed $EB$ angular power spectrum.

An identical conclusion is easily derived for the case of the lensed $EE$ and $BB$ spectrum as the primary $EB$ power spectrum does not contribute to the nesed $EE$ and $BB$ spectra, since any contribution of $C^{EB}_\ell$ to $\tilde{C}^{EE(BB)}_\ell$ arises as either proportionnal to ${}_{(+)}I^{\ell\ell_1\ell_2}_{mm_1m_2}{}_{(-)}I^{\ell\ell_1\ell_2~\star}_{mm_1m_2}$ or proportionnal to $\sum_{m_3}{}_{(-)}J^{\ell\ell_3\ell\ell_3}_{mm_3mm_3}$, which are both vanishing. One finally obtains the following expressions for the lensed angular power spectra
\begin{eqnarray}
	\tilde{C}^{EE}_\ell&=&\left[1+R^{P}\right]C^{EE}_\ell+\displaystyle\sum_{\ell_1,\ell_2}F^{(+)}_{\ell\ell_1\ell_2}C^{\phi\phi}_{\ell_1}C^{EE}_{\ell_2} \\
	&+&\displaystyle\sum_{\ell_1,\ell_2}F^{(-)}_{\ell\ell_1\ell_2}C^{\phi\phi}_{\ell_1}C^{BB}_{\ell_2} \nonumber \\
	\tilde{C}^{BB}_\ell&=&\left[1+R^{P}\right]C^{BB}_\ell+\displaystyle\sum_{\ell_1,\ell_2}F^{(+)}_{\ell\ell_1\ell_2}C^{\phi\phi}_{\ell_1}C^{BB}_{\ell_2} \\
	&+&\displaystyle\sum_{\ell_1,\ell_2}F^{(-)}_{\ell\ell_1\ell_2}C^{\phi\phi}_{\ell_1}C^{EE}_{\ell_2} \nonumber \\
	\tilde{C}^{EB}_\ell&=&\left[1+R^{P}\right]C^{EB}_\ell \\
	&+&\displaystyle\sum_{\ell_1,\ell_2}\left(F^{(+)}_{\ell\ell_1\ell_2}-F^{(-)}_{\ell\ell_1\ell_2}\right)C^{\phi\phi}_{\ell_1}C^{EB}_{\ell_2}, \nonumber
\end{eqnarray}
with
\begin{eqnarray}
	F^{(\pm)}_{\ell\ell_1\ell_2}&=&\frac{1}{4}\displaystyle\sum_{m_1,m_2}\left|{}_{(\pm)}I^{\ell\ell_1\ell_2}_{mm_1m_2}\right|^2, \\
	R^P&=&\frac{1}{2}\displaystyle\sum_{\ell_3}C^{\phi\phi}_{\ell_3}\sum_{m_3}\mathrm{Re}\left[{}_{(+)}J^{\ell\ell_3\ell\ell_3}_{mm_3mm_3}\right].
\end{eqnarray}
The computation of $R^P$ directly follows from the computation of $R^{TB}$ and gives
\begin{equation}
	R^P=-\frac{1}{2}\left[\ell(\ell+1)-4\right]\displaystyle\sum_{\ell_3}\ell_3(\ell_3+1)\frac{2\ell_3+1}{4\pi}C^{\phi\phi}_{\ell_3}.
\end{equation}
Using the expressions of ${}_{(\pm)}I^{\ell\ell_1\ell_2}_{mm_1m_2}$ as functions of the Wigner-$3j$'s and the summation rule for the product of two such symbols, one easily derive that
\begin{eqnarray}
	F^{(\pm)}_{\ell\ell_1\ell_2}&=&\frac{1}{16}\left[\ell_1(\ell_1+1)+\ell_2(\ell_2+1)-\ell(\ell+1)\right]^2  \nonumber \\
	&\times&{\frac{(2\ell_1+1)(2\ell_2+1)}{4\pi}} \\
	&\times&\left[\left(\begin{array}{ccc}
			\ell & \ell_1 & \ell_2 \\
			2 & 0 &-2
		\end{array}\right)\pm\left(\begin{array}{ccc}
			\ell & \ell_1 & \ell_2 \\
			-2 & 0 &2
		\end{array}\right)\right]^2. \nonumber
\end{eqnarray}

\section{Fisher matrix and miscalibration angle}
\label{appb}
Let us consider the simple situation where the mode-counting derivation of the covariance matrix on the angular power spectra is used. In that case, the Fisher information matrix can be re-expressed as 
\begin{equation}
	\left[\mathbf{F}\right]_{ij}=\frac{f_\mathrm{sky}}{2}\ds\sum_{\ell,m}\mathrm{Tr}\left[\frac{\partial\mathbf{C}^{\mathrm{(obs)}}_{\ell m}}{\partial\theta_i}{\mathbf{C}^{\mathrm{(obs)}}_{\ell m}}^{-1}\frac{\partial\mathbf{C}^{\mathrm{(obs)}}_{\ell m}}{\partial\theta_j}{\mathbf{C}^{\mathrm{(obs)}}_{\ell m}}^{-1}\right],
\end{equation}
with $\mathbf{C}^{\mathrm{(obs)}}_{\ell m}=\left<\mathbf{a}^{\mathrm{(obs)}}_{\ell m}\mathbf{a}^{\mathrm{(obs)}~\dag}_{\ell m}\right>$ the covariance matrix of the $T,~E$ and $B$ multipoles of the observed CMB maps. (We remind that in the mode-counting approximation, those covariance matrices are supposed to be block diagonal in $(\ell,m)$-space.) If those multipoles are affected by a global miscalibration, they are related to the CMB multipoles by a rotation matrix :
\begin{equation}
	\mathbf{a}^{\mathrm{(obs)}}_{\ell m}=\mathbf{R}(2\Delta\psi)\times\mathbf{a}_{\ell m}+\mathbf{n}_{\ell m},
\end{equation}
with	
\begin{equation}
	\mathbf{a}^{\mathrm{(obs)}}_{\ell m}=\left(\begin{array}{c}
		a^{T~\mathrm{(obs)}}_{\ell m} \\
		a^{E~\mathrm{(obs)}}_{\ell m} \\
		a^{B~\mathrm{(obs)}}_{\ell m}
	\end{array}\right)
\end{equation}
the observed (systematically rotated and noisy) multipoles,
\begin{equation}
	\mathbf{a}_{\ell m}=\left(\begin{array}{c}
		a^{T}_{\ell m} \\
		a^{E}_{\ell m} \\
		a^{B}_{\ell m}
	\end{array}\right)
\end{equation}
the intrinsic (noiseless and unrotated) CMB multipoles, 
\begin{equation}
	\mathbf{n}_{\ell m}=\left(\begin{array}{c}
		n^{T}_{\ell m} \\
		n^{E}_{\ell m} \\
		n^{B}_{\ell m}
	\end{array}\right)
\end{equation}
the instrumental noise, and, 
\begin{equation}
	\mathbf{R}(2\Delta\psi)=\left(\begin{array}{ccc}
		1 & 0 & 0 \\
		0 & \cos(2\Delta\psi) & -\sin(2\Delta\psi) \\
		0 & \sin(2\Delta\psi) & \cos(2\Delta\psi)
	\end{array}\right)
\end{equation}
a rotation matrix modeling the impact of miscalibration. Being a rotation matrix, it satisfies $\mathbf{R}^\dag(2\Delta\psi)=\mathbf{R}(-2\Delta\psi)=\mathbf{R}^{-1}(2\Delta\psi)$.

On defining $\mathbf{C}_{\ell m}=\left<\mathbf{a}_{\ell m}\mathbf{a}^{\dag}_{\ell m}\right>$ and $\mathbf{N}_{\ell m}=\left<\mathbf{n}_{\ell m}\mathbf{n}^{\dag}_{\ell m}\right>$, both assumed to be block diagonal -- which is only true for homogeneous noise --, it is easily shown that
\begin{equation}
	\mathbf{C}^{\mathrm{(obs)}}_{\ell m}=\mathbf{R}(2\Delta\psi)\mathbf{C}_{\ell m}\mathbf{R}^{-1}(2\Delta\psi)+\mathbf{N}_{\ell m}.
\end{equation}
Assuming finally that we are in such an experimental setups where sampling variance is dominating, then one can neglect noise leading to $\mathbf{C}^{\mathrm{(obs)}}_{\ell m}=\mathbf{R}\mathbf{C}_{\ell m}\mathbf{R}^{-1}$ and ${\mathbf{C}^{\mathrm{(obs)}}_{\ell m}}^{-1}=\mathbf{R}{\mathbf{C}^{-1}_{\ell m}}\mathbf{R}^{-1}$. 

It is straighforward to show that the Fisher matrix for $r_{(\pm)}$ reduces to:
\begin{equation}
	\left[\mathbf{F}\right]_{r_{(\pm)}r_{(\pm)}}=\frac{f_\mathrm{sky}}{2}\ds\sum_{\ell,m}\mathrm{Tr}\left[\frac{\partial\mathbf{C}_{\ell m}}{\partial r_{(\pm)}}{\mathbf{C}^{-1}_{\ell m}}\frac{\partial\mathbf{C}_{\ell m}}{\partial r_{(\pm)}}{\mathbf{C}^{-1}_{\ell m}}\right].
\end{equation}
This simply means that for an experiment only limited by cosmic variance, the block $r_{(\pm)}$ of the Fisher information matrix is exactly the same as if there were no rotation.

Let us now consider the off-diagonal terms amounting for the correlations between $r_{(\pm)}$ and $\Delta\psi$, {\it i.e.}
\begin{eqnarray}
	\left[\mathbf{F}\right]_{r_{(\pm)}\Delta\psi}&=&\frac{f_\mathrm{sky}}{2}\ds\sum_{\ell,m}\mathrm{Tr}\left[\left(\frac{\partial \mathbf{R}}{\partial\Delta\psi}\mathbf{C}_{\ell m}\mathbf{R}^{-1}+\mathbf{R}\mathbf{C}_{\ell m}\frac{\partial\mathbf{R}^{-1}}{\partial\Delta\psi}\right)\right. \nonumber \\
	&\times&\left.\mathbf{R}\mathbf{C}^{-1}_{\ell m}\frac{\partial\mathbf{C}_{\ell m}}{\partial r_{(\pm)}}\mathbf{C}^{-1}_{\ell m}\mathbf{R}^{-1}\right].
\end{eqnarray}
Using the fact we are considering the trace and $\mathbf{M}^{-1}\left(\partial\mathbf{M}\right)+\left(\partial\mathbf{M}^{-1}\right)\mathbf{M}=0$, the above expression gives:
\begin{eqnarray}
	\left[\mathbf{F}\right]_{r_{(\pm)}\Delta\psi}&=&\frac{f_\mathrm{sky}}{2}\ds\sum_{\ell,m}\mathrm{Tr}\left[\mathbf{R}^{-1}\frac{\partial \mathbf{R}}{\partial\Delta\psi}\right. \\
	&\times&\left.\left(\frac{\partial\mathbf{C}_{\ell m}}{\partial r_{(\pm)}}\mathbf{C}^{-1}_{\ell m}-\mathbf{C}^{-1}_{\ell m}\frac{\partial\mathbf{C}_{\ell m}}{\partial r_{(\pm)}}\right)\right], \nonumber
\end{eqnarray}	
with
\begin{equation}
	\mathbf{R}^{-1}\frac{\partial \mathbf{R}}{\partial\Delta\psi}=\left(\begin{array}{ccc}
		0 & 0 & 0 \\
		0 & 0 & -2 \\
		0 & 2 & 0
	\end{array}\right).
\end{equation}
Because $\left(\frac{\partial\mathbf{C}_{\ell m}}{\partial r_{(\pm)}}\mathbf{C}^{-1}_{\ell m}-\mathbf{C}^{-1}_{\ell m}\frac{\partial\mathbf{C}_{\ell m}}{\partial r_{(\pm)}}\right)$ does not depend on $\Delta\psi$, this means that the correlations between parameters $r_{(\pm)}$ and $\Delta\psi$ are independent of the value of $\Delta\psi$. 

\end{appendix}


\end{document}